\documentclass[onecolumn]{aastex631} 

\usepackage{natbib}

\shorttitle{ATLAS}
\shortauthors{Kar et al.}

\begin{document}

\title{The Solar Neighborhood LI: A Variability Survey of Nearby M Dwarfs with Planets from Months to Decades with \textit{TESS} and the CTIO/SMARTS 0.9 m Telescope}

\author[0000-0002-9811-5521]{Aman Kar}
\affiliation{Department of Physics and Astronomy, Georgia State University, Atlanta, GA 30303, USA}
\affiliation{RECONS Institute, Chambersburg, PA 17201, USA}

\author[0000-0002-9061-2865]{Todd J. Henry}
\affiliation{RECONS Institute, Chambersburg, PA 17201, USA}

\author[0000-0001-9834-5792]{Andrew A. Couperus}
\affiliation{Department of Physics and Astronomy, Georgia State University, Atlanta, GA 30303, USA}
\affiliation{RECONS Institute, Chambersburg, PA 17201, USA}

\author[0000-0002-1864-6120]{Eliot Halley Vrijmoet}
\affiliation{Five College Astronomy Department, Smith College, Northampton, MA 01063, USA}
\affiliation{RECONS Institute, Chambersburg, PA 17201, USA}

\author[0000-0003-0193-2187]{Wei-Chun Jao}
\affiliation{Department of Physics and Astronomy, Georgia State University, Atlanta, GA 30303, USA}

\received{17 November 2023}
\revised{3 February 2024}
\accepted{16 February 2024}

\begin{abstract}

We present the optical photometric variability of 32 planet-hosting M dwarfs within 25 parsecs over timescales of months to decades. The primary goal of this project, ATLAS --- A Trail to Life Around Stars, is to follow the trail to life by revealing nearby M dwarfs with planets that are also ``quiet", which may make them more amiable to habitability. There are 69 reported exoplanets orbiting the 32 stars discussed here, providing a rich sample of worlds for which environmental evaluations are needed. We examine the optical flux environments of these planets over month-long timescales for 23 stars observed by {\it TESS}, and find that 17 vary by less than 1\% ($\sim$11 mmag). All 32 stars are being observed at the CTIO/SMARTS 0.9 m, with a median duration of 19.1 years of optical photometric data in the $VRI$ bands. We find over these extended timescales that six stars show optical flux variations less than 2\%, 25 vary from 2--6\% ($\sim$22-67 mmag), and only one, Proxima Centauri, varies by more than 6\%. Overall, LHS 1678 exhibits the lowest optical variability levels measured over all timescales examined, thereby providing one of the most stable photometric environments among planets reported around M dwarfs within 25 parsecs. More than 600 of the nearest M dwarfs are being observed at the 0.9 m in the RECONS program that began in 1999, and many more planet hosts will undoubtedly be revealed, providing more destinations to be added to the ATLAS sample in the future.

\end{abstract}


\section{Introduction} \label{Intro}

M dwarfs are the most common type of stars in the solar neighborhood \citep{Henry2006,Henry2018} and presumably throughout the Milky Way and other galaxies. They represent 75\% of the stars in the solar neighborhood and in fact, provide more aggregate habitable zone (HZ) real estate than any other stellar type \citep{Cantrell2013}, and have been found to have closely-packed sets of terrestrial planets \citep{Shields2016}. They are cooler and dimmer than more massive stars and consume their hydrogen slowly over extraordinarily long timescales, creating enduring stable environments in which life might originate and thrive. Pragmatically, M dwarfs are excellent candidates to search for other worlds because their small stellar radii and masses permit the detection of Earth-size planets, which are anticipated to be common \citep{Dressing&Charbonneau2015}. 

With the discovery of thousands of exoplanets, the field of exoplanetary science has rapidly developed in the last few decades, thanks to space missions like \textit{Kepler, K2,} and the ongoing Transiting Exoplanet Survey Satellite (\textit{TESS}) effort. Our closest neighbor, the M5.0V star Proxima Centauri, has been reported to host two or three planets \citep{Anglada2016,Damasso2020,Mascareno2020,Artigau2022}, and dozens of other nearby M dwarfs are reported to be orbited by exoplanets, typically terrestrial in nature. 

Although M dwarfs are often known to be flare stars, such outbursts are not necessarily unfavorable for the habitability of orbiting planets because most of the (presumed) life-damaging UV radiation affects only the stratosphere where ozone is photolyzed, and thus does not reach the surface of the planet \citep{Tarter2007,Segura2010}. Still, frequent stellar activity might damage a planetary atmosphere irreparably, or reduce it to a level from which it may not recover fast enough for life to endure \citep{Tilley2019}. Thus, such flaring events may play a key role in the habitability of planets around the host star, although \citet{Ilin2021} found that giant flares tend to occur at higher latitudes for fully convective, late-type M dwarfs, which could minimize the impact of flares on planets orbiting these stars' equatorial regions. The history, duration, and location of activity and flares all need to be probed to understand their effects on the atmospheres of exoplanets orbiting M dwarfs. Current activity levels may provide information about the age of a star because young stars are known to be active, while the absence of an atmosphere around an exoplanet may indicate past activity levels of the host star. Also, prolonged periods of low stellar activity observed over several years may suggest long-term stability, as opposed to random observations during the minimum of the stellar activity cycle when observed over shorter timescales. Thus, among M dwarfs, those with minimal stellar activity likely provide better, or a less worse, environments for life on an orbiting planet because lower levels of stellar activity may allow an atmosphere to be chemically stable and preserved.

The REsearch Consortium On Nearby Stars (RECONS, \textit{www.recons.org}) is a multi-decade effort to discover and characterize members of the solar neighborhood \citep{Henry1997,Jao2005}. One aspect of the RECONS effort is an observing program to secure long-term astrometric and photometric data on stars within 25 pc, with a current focus on a sample of 611 M dwarfs targeted with the Small and Moderate Aperture Research Telescope System (SMARTS) 0.9 m telescope at the Cerro Tololo Inter-American Observatory (CTIO) in Chile. In this paper we describe results of optical photometric variability for 32 M dwarfs within 25 pc on the 0.9 m program that have 69 reported planets, some of which have been observed for more than 20 years. Previous variability studies from this effort have revealed a decrease in observed variability levels of M dwarfs from bluer to redder wavelengths \citep{Hosey2015}, and high variability levels for M dwarfs above the main sequence \citep{Clements2017}. This paper builds on those previous studies by investigating M dwarf exoplanet hosts to identify stars that show the least photometric variability, potentially offering the most stable environments and consequently providing the best chances for life-bearing worlds. In $\S$\ref{ATLAS} we give an overview of the ATLAS project, followed in $\S$\ref{Sample} by a description of the sample and selection criteria. In $\S$\ref{RECONS} the RECONS data and long-term variability results from the SMARTS 0.9 m telescope are described. These long-term results are augmented with mid-term variability results from \textit{TESS} in $\S$\ref{TESS}. We discuss the long- and mid-term results in $\S$\ref{Discussion}, provide details about systems worthy of note in $\S$\ref{Systems}, and outline our conclusions and future work in $\S$\ref{Conclusion}.

\section{The ATLAS Project} \label{ATLAS}

ATLAS (A Trail to Life Around Stars) is the project described here, with an aim to find stellar systems with the most habitable environments in the solar neighborhood, defined here to be within 25 pc. Habitability is an essential factor in gauging the importance of a particular planetary system. Traditionally, a planet's habitability has been defined in terms of the irradiation it receives from its host star, given its orbital distance and the potential for liquid water on its surface. However, the habitability of a system may depend on a vast range of parameters of both the star and the planet, ranging from planet-star tidal interaction \citep{Griebmeier2005,Barnes2008,Barnes2013,Jackson2008} to geologically sustainable habitability \citep{Kasting1993,Williams1997,Gaidos2005,Scalo2007,Foley2015}. In this paper, we focus on the stellar activity of the host star as a relevant factor in the habitability of a planetary environment, where changes in stellar flux levels define activity. Causes of this stellar activity can be categorized into three distinct variability regimes: short-term variations lasting minutes to hours due to stellar flares, mid-term variations from days to months caused by stellar rotation, and long-term variations stretching from years to decades manifested by stellar cycles.

Tracers of stellar activity at various wavelengths probe different layers of a star. Coronal activity can be traced by monitoring the ultraviolet and x-ray fluxes from an M dwarf, where fast rotators with $P_{rot} <$ 10 days show elevated levels of high energy fluxes compared to slow rotators \citep{Magaudda2020}. Chromospheric activity is commonly traced by the H$\alpha$ emission line, a diagnostic tool that can differentiate between active and inactive M dwarfs \citep{Newton2017}, although the precise methods of defining ``active'' vs.~``inactive'' stars vary. The photosphere is evident at optical wavelengths, where changes in flux levels correspond to starspots coming in and out of view due to stellar rotation, and over longer timescales changes occur when spot numbers and coverage fractions potentially evolve over time. Previous studies have found that M dwarfs vary in the optical due to flares \citep{Segura2010,Davenport2012} and stellar rotation \citep{McQuillan2014,Reinhold2020,Lu2022}, at least the latter of which is correlated to chromospheric activity \citep{Mohanty2003}. It has also been shown that fast-rotating M dwarfs typically have higher amplitudes of optical variability than slow rotators \citep{McQuillan2014}, and in extreme cases for young M dwarfs, rotation modulations can be as high as 25\% at optical wavelengths \citep{Rodono1986, Messina2003}. What has yet to be investigated thoroughly are the photospheric changes over years to decades, a timescale we begin to examine in this paper. 

Every indicator of variability can be considered in terms of its timescale, with the most commonly traced activity being photospheric starspot variations. The \textit{TESS} mission offers high cadence coverage over about a month per visit, enabling the determination of the mid-term variability of an M dwarf. The detection of multi-year stellar cycles, however, requires long-term monitoring, and the RECONS effort is one of the few long-term surveys where variability due to starspot cycles can be observed. Spot cycles \citep{Gomes2011,Suarez2016} occur at longer timescales than flares and rotation --- one clear example is our Sun, on which the number of sunspots changes over its 11-year activity cycle, with more sunspots observed during solar maxima and few to no sunspots during solar minima \citep{Balogh2014}. It is important to note that several studies have found that the change in optical light fluxes due to stellar cycles is measurable, but low, because the amplitude of the overall variability is often $\lesssim$ 2\% for late-type field stars \citep{Hosey2015,Mascareno2016,Mignon2023}. M dwarfs have been found to display stellar activity cycles on the order of at least several years, e.g., \citet{Cincunegui2007,Buccino2011,Gomes2012,Robertson2013,Hosey2015,Clements2017,Henry2018}. Cycles reported in these references and to be published from our long-term program span a large range in duration, from a bit over a year to at least a few decades.

In this work ($\S$\ref{Discussion}) we show that for nearby M dwarfs that are presumably older than 1 Gyr, the optical variability can be up to $\sim$8\% over long timescales. In this first installment of the ATLAS project, we evaluate the mid-term and long-term variability of 32 M dwarfs within 25 pc reported to have planets, with the goal of identifying the stars exhibiting the least variability, and which potentially offer the most stable flux environments where life could thrive on the orbiting planets.

\vskip 30pt
\section{Sample} \label{Sample}

\begin{figure}
\plotone{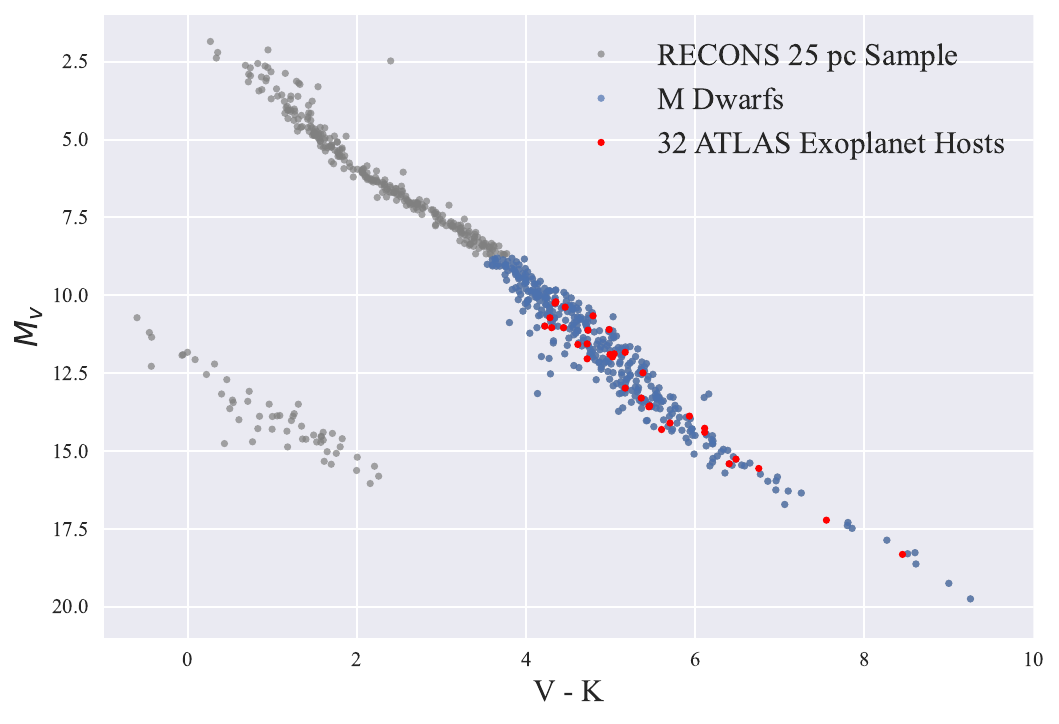}
\caption{Hertzsprung-Russell diagram highlighting the 32 ATLAS exoplanet hosts (red) plotted with the RECONS 25 pc sample (grey) and M dwarfs (blue), with limits in $M_V$ and $V-K$ set as described in $\S$\ref{Sample}.}
\label{fig:HRD}
\end{figure}

In the RECONS sample of stellar systems within 25 parsecs of the Sun, there are $\sim$3000 containing at least one M dwarf. For this survey, our sample consists of systems within this horizon that contain an M dwarf and at least one confirmed exoplanet. M dwarfs were selected using a combination of $V$-band absolute magnitude through a range of $8.8 \leq M_V \leq 20.0$ and within a $V-K$ color range of $3.7 \leq V-K \leq 9.5$, limits derived using the \citet{Benedict2016} $V$-band mass-luminosity relation for main sequence M dwarfs. The $V$ photometry is primarily from our CTIO/SMARTS 0.9 m telescope program, while the $K$ photometry is extracted from the 2MASS catalog. These magnitude limits and color cutoffs correspond to mass limits of 0.075 $\leq M/M_{\odot} \leq$ 0.63. We then trimmed this list to include stars with declinations from +30$^{\circ}$ to $-$90$^{\circ}$ and with $V >$ 10 because these are targeted in the long-term RECONS astrometric program.

This process yielded a list of nearby M dwarfs that we crossmatched with the NASA Exoplanet Archive\footnote{https://exoplanetarchive.ipac.caltech.edu/} in January 2023, when it listed 5235 exoplanets, including 2710 discoveries made by \textit{Kepler}, 543 by \textit{K2}, and 285 by \textit{TESS}. These systems were checked against the \textit{Gaia} DR3 results \citep{Gaia2016,GaiaDR32023} for \textit{Gaia} trigonometric parallax $\pi \geq$ 40 mas. We refined our list to include only systems in which at least one exoplanet was reported to orbit an M dwarf component, i.e., our list includes multiple star systems in which there may be larger stars such as $\alpha$ Centauri A and B plus Proxima system. 

Our selection criteria resulted in 32 systems with 69 reported exoplanets that have at least three years of RECONS observations in the ongoing program \citep{Henry2018}. These 32 M dwarfs that constitute the ATLAS sample are shown in the observational HR diagram of Figure \ref{fig:HRD}, are listed in Table \ref{tab:ATLAS32}, and comprise the set for which we discuss our variability measurements from RECONS and \textit{TESS} observations in the following sections.

\section{RECONS Long-term Data AND Results} \label{RECONS}

Stellar cycles analogous to the 11-year solar cycle may play important roles in planetary habitability. Compared to the more often studied rotation variability changes that occur over hours to months, stellar cycles occur over years and require long-term monitoring efforts to be characterized. We used RECONS data from the CTIO/SMARTS 0.9 m telescope for our long-term variability study of the 32 ATLAS catalogue stars described here.

\subsection{The CTIO/SMARTS 0.9 m Telescope and Camera} \label{0.9m Camera}

Since 1999, the RECONS program has conducted astrometric and photometric measurements of red dwarfs in the solar neighborhood with the CTIO/SMARTS 0.9 m telescope \citep{Jao2005,Henry2018}. The camera mounted on the telescope has a 2048 $\times$ 2046 Tektronix CCD with a 0\farcs401 pixel$^{-1}$ plate scale, resulting in a 13\farcm6 $\times$ 13\farcm6 field of view (FOV). The program uses the center quarter of the chip, with a 6\farcm8 $\times$ 6\farcm8 FOV, as this improves the astrometry and resulting parallax measurements \citep{Jao2005}.  A set of 5\textendash15 reference stars within this FOV are used for both the astrometry and variability measurements via differential measurements. Observations are made using Johnson-Kron-Cousins \textit{VRI} filters with central wavelengths of 5438 \AA, 6425 \AA, and 8075 {\AA}, respectively \citep{Jao2011}. In 2005 March, the `old' Tek \#2 \textit{V} filter was swapped with a similar `new' Tek \#1 \textit{V} filter because the former cracked in the corner and the latter, with a central wavelength of 5475 \AA, was used until 2009 June when the `old' filter was reinstituted for observations. These filter changes caused astrometric shifts during this period relative to previous data, but no significant photometric offsets were measureable, so the switched filter interlude is of no concern for our variability study (for more details, see \citet{Subasavage2009} and \citet{Riedel2010}). 
 
\subsection{Observations for the RECONS Long-term Program} \label{0.9m Obs}

Observing stars in the RECONS program uses well-honed protocols that ensure the data quality is consistent over varying seasons, sky conditions, and observers over the years. Each RECONS target is visited at least twice a year, with five frames typically taken per visit, resulting in at least 10 frames every year for each target. The target is placed on the CCD such that the set of 5\textendash15 field stars are positioned within the frame, typically within a few pixels on the chip for every epoch. Exposure times are scaled by the brightnesses of the target and field stars, and usually range from 30\textendash300 seconds. These exposure times are adjusted frame to frame to accommodate changes in seeing, cloud coverage, and to ensure the target star is not saturated on the CCD. We typically expose until the target star reaches $\sim$50,000 peak counts to achieve a S/N of $>$ 100, which usually ensures the reference stars have at least $\sim$10,000 peak counts as well, although that is not always possible. Frames are taken within 120 minutes of the target transiting the meridian to minimize corrections required for differential color refraction for the astrometry aspect of the scientific effort.

\subsection {Photometric Reductions for the RECONS Long-term Data} 
\label{0.9m Data Reduction}

The RECONS data reduction methodology is described in detail in \citet{Jao2005,Henry2006,Winters2011}. The reduction techniques are briefly captured in the following steps: (1) Typically, calibration frames are taken for flat-fielding and bias subtraction at the beginning of each night. These corrections are later performed with standard IRAF routines that produce our calibrated science frames. (2) Each calibrated science frame is tagged for the target star and reference stars in the ensemble of 5\textendash15 field stars. We also check for saturation of these stars and discard frames accordingly. (3) To compute instrumental magnitudes of the target and reference stars for each frame, a circular Gaussian profile is scaled to the light distribution of each source, and the source pixel values within this Gaussian window are integrated (MAG\_WIN parameter via SExtractor) \citep{Bertin1996}. This process also determines the centroids of each star in each frame. (4) These instrumental magnitudes still need to be corrected because they contain offsets resulting from different sky conditions, airmasses, and exposure times for each frame. This correction is achieved by performing relative photometry following a prescription from \citet{Honeycutt1992} where the deviations of all available reference stars in all frames are simultaneously minimized to determine corrective offsets for each frame, with the Gaussfit program \citep{Jefferys1987} used to carry out the least-squares minimization. We discard any highly variable reference star at this step. (5) Our final corrected instrumental magnitudes are then utilized to calculate the variability of the target star for this study, where nightly means of frames taken typically within 30 minutes are used rather than individual frame values.

\subsection{Variability Characterization} \label{Variability}

The characterization of variability in a star depends on the timescale, wavelength, and tools used to quantify signal variations. The standard deviation, $\sigma$ is often used to characterize the photometric variability of a given star from time-series data \citep[e.g.,][]{Jao2011, Hosey2015}. This quantity captures the dispersion of data points from the mean, but can be skewed by data value outliers, and also assumes that the data values follow a normal distribution. As outlined below, a Gaussian often does not describe the full distribution of these data well, especially when only a partial rotation or stellar cycle period is covered.

\begin{figure}
\epsscale{1.16}
\plottwo{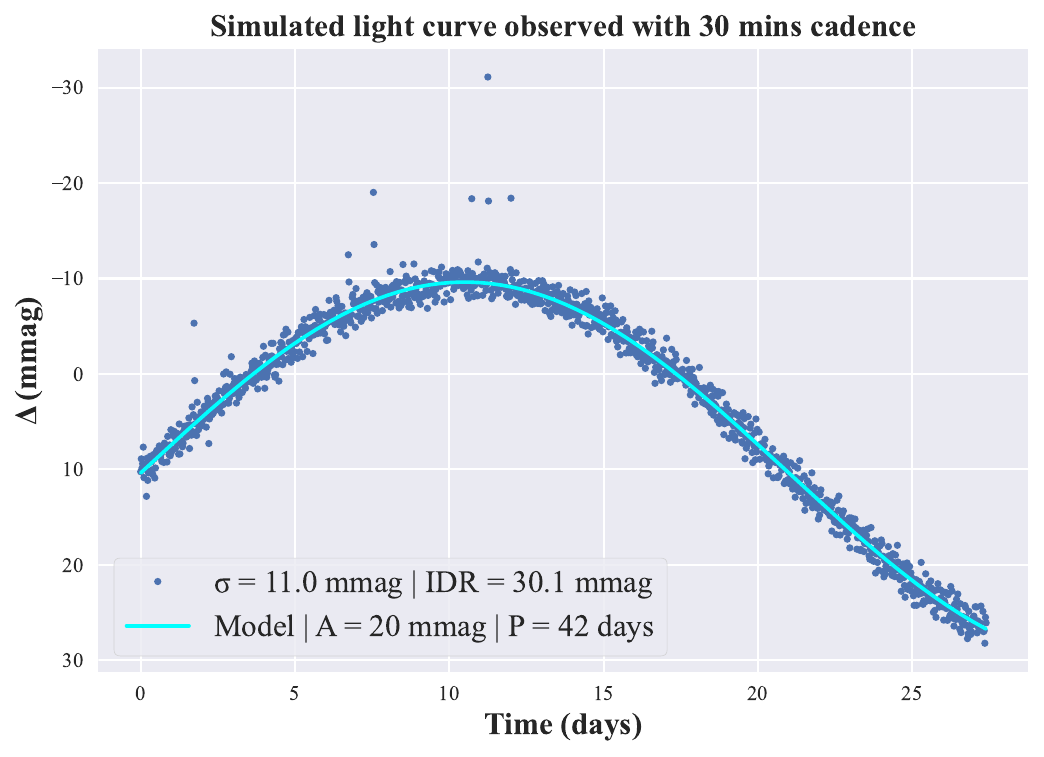}{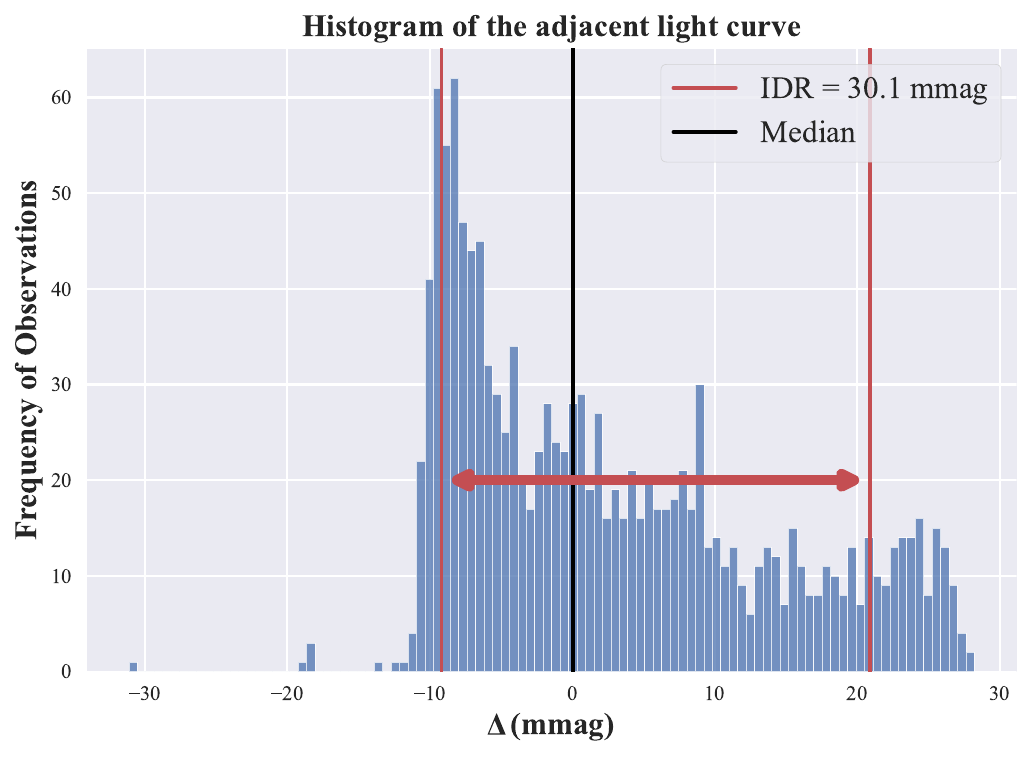}
\caption{\textbf{left}: Simulated light curve of a star with a 42-day rotation period and a semi-amplitude of 20 mmag over a baseline of 27.4 days. Each simulated $y$ value (blue) is plotted along the $y - Z$ sine curve (cyan; see Eq. 1). The zero y-axis value corresponds to the median magnitude of the light curve. \textbf{right}: A histogram of the simulated light curve is given with the median value represented with a black vertical line, highlighting the 50th percentile of the distribution. The Interdecile Range (red horizontal arrows) used in this paper ranges from the 90th percentile to the 10th percentile (red vertical lines), and better represents the overall range of variability in this example.}
\label{fig:sim}
\end{figure}

To explore the measurement of variability, in Figure \ref{fig:sim} (left) we simulated a star's light curve with a rotation period of 42 days and a semi-amplitude of 20 mmag. This is representative of a \textit{TESS} observation, but can apply to longer timescales such as the RECONS datasets discussed here as well; it is simply a matter of changing the units on the time axis in a plot. We modeled this light curve using a simple sine curve of the form: 

\begin{equation}\label{eq:1}
 y = A \sin \frac{2\pi x}{P} + Z
\end{equation}

\noindent where $A$ is the semi-amplitude, $P$ is the period, $Z$ is the additive white noise, and $x$ is the instance in time where $y$ is calculated. Our observation baseline is set to 27.4 days to mimic the duration of the observing window for a sector by \textit{TESS}. We model the light curve at a cadence the same as \textit{TESS}, which is one frame every 30 minutes. We also randomly inject flares, so the simulated light curve better represents an M dwarf that is at least moderately active; the light curve in Figure \ref{fig:sim} (left) shows a few of these flares. Figure \ref{fig:sim} (right) shows a histogram of the light curve, and it is immediately obvious that the underlying distribution is not Gaussian in shape. We conclude that although standard deviation is a standard method to quantify variability, it does not typically represent the spread of the data in this application. Other metrics that are proxies for magnetic activity such as $R_{var}$ \citep{Basri2013} and $\langle S_{ph} \rangle$ \citep{Mathur2014} have also been used. $R_{var}$ is calculated as the difference between the 95th and 5th percentile of the flux distribution over one rotational timescale while $\langle S_{ph} \rangle$ is the mean value of the standard deviations over a time interval of five times the rotational period of the star. However, both of these metrics are more accurate when the observations are evenly sampled and at least a full rotation period of the star is covered, attributes often not characteristic of the data sets used here.

Instead, we use the Interdecile range (IDR) that characterizes variability by measuring the dispersion capturing 80\% of the time series data centered about the median. This is calculated by measuring the difference between the 90th and 10th percentiles of the distribution. We avoid using the entire range because just a single outlier (e.g., caused by a flare) would expand the calculated range beyond our goal of understanding the typical range of brightness levels in the targeted M dwarfs. Because the activity period (due to rotation or cycles) of our stars is not always known, we measure this over the entire set of time series data available. For stars with activity periods less than the observation baselines, IDR represents the majority of the variability in the given time series data, while for stars with longer activity periods, IDR allows us to set the minimum value of the variability. We adopt IDR as our variability measurement tool because it is more robust to outliers and can better characterize the overall spread of the data than the standard deviation. For example, for the simulated stellar light curve in Figure \ref{fig:sim}, we find a $\sigma$ of 11.0 mmag and an IDR of 30.1 mmag. In this paper, we report variability measurement amplitudes for our ATLAS sample with IDR for both the long-term cycle data and the mid-term rotation results. For the long-term observations, the IDR for the variability noise floor is set at 10 mmag and determined from observations of non-varying photometric white dwarfs \citep{Jao2011}.

\subsection{Results from the RECONS Long-term Data} \label{0.9m Results}

\begin{figure}
\epsscale{1.1}
\plotone{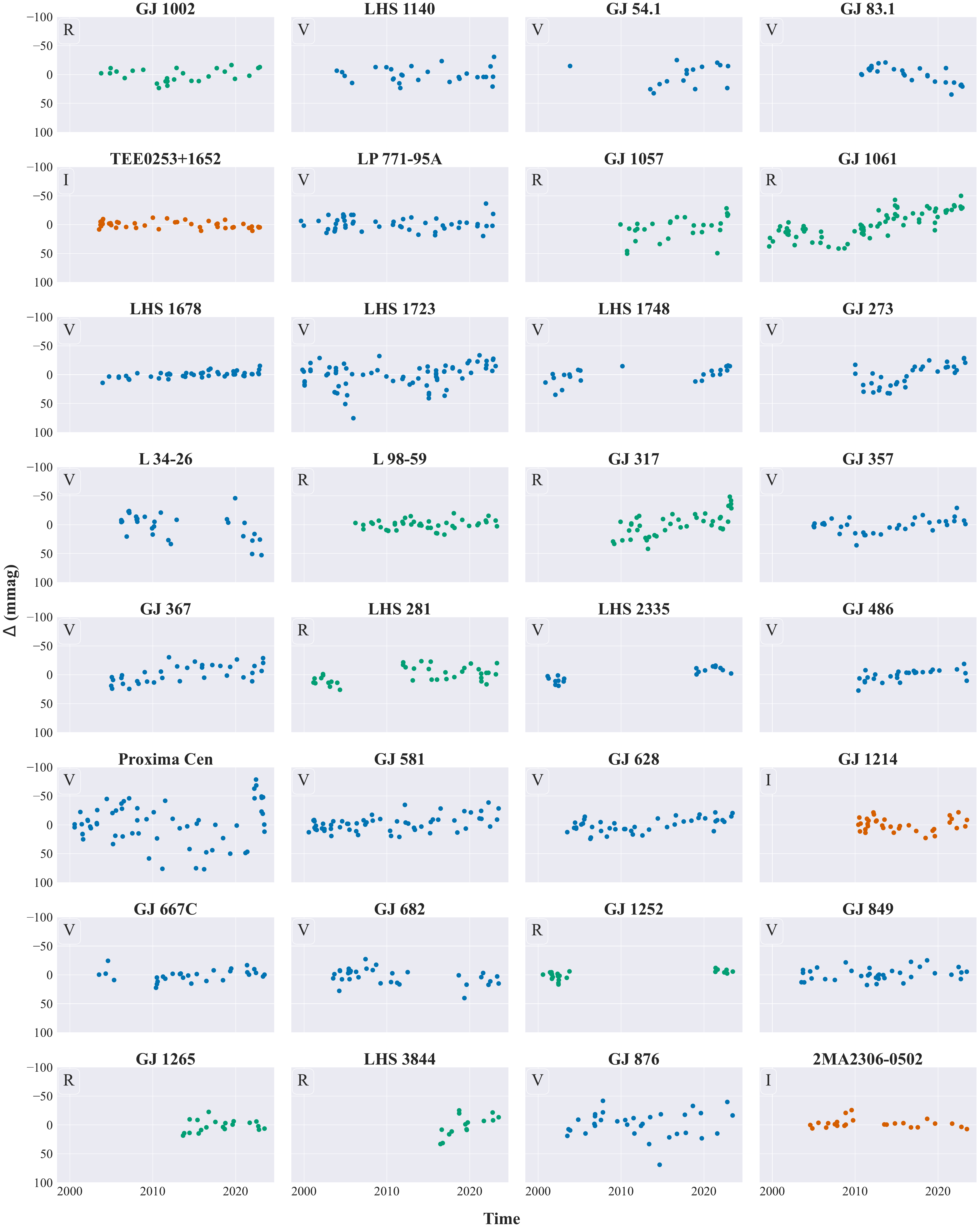}
\caption{Light curves of 32 ATLAS targets obtained using data from the CTIO/SMARTS 0.9 m telescope. Plots are ordered by Right Ascension from left to right and top to bottom. Blue, green, and orange points represent observations taken in filter \textit{V}, \textit{R}, and \textit{I}, respectively. IDR measurements are reported from all the light curves in Table \ref{tab:ATLAS32}. Each data point represents the nightly mean of observations, typically taken within a window of 30 minutes. The zero y-axis value corresponds to the median magnitude of each system.}
\label{fig:RECONS}
\end{figure}

We have performed a complete set of new reductions for the 32 ATLAS stars following the data reduction method outlined in $\S$\ref{0.9m Data Reduction}; values reported here supersede those given in previous papers in this series. The variability results from the RECONS 0.9 m program are given in Table \ref{tab:ATLAS32}, where the first column group lists ATLAS star names (1), followed by J2000 Right Ascension (2) and Declination (3) \textit{Gaia} DR3 coordinates. The following seven columns (4--10) are the RECONS 0.9 m results, including apparent \textit{VRI} magnitudes (4--6), the filter used for the set of observations (7), the time coverage in years (8), the standard deviation $\sigma$ (9) for comparison to previous and others' efforts, and the IDR (10) range in the light curve. The remaining columns (11--17) relate to \textit{TESS} and are discussed below in $\S$\ref{TESS Results}. The \textit{VRI} magnitudes have been measured at the 0.9 m by observing target stars and standard stars on photometric nights (for more details, see \citealt{Hosey2015}). Light curves for all 32 ATLAS targets are given in Figure \ref{fig:RECONS}. Our observations span 1999 to 2023 with a median coverage of 19.1 years for the 32 target stars. Each panel represents the long-term light curve for an M dwarf, with remarkable ranges between minimally variable stars like LHS 1678 and highly variable stars like Proxima Cen. Roughly half of the stars show consistent data sets over 20 years, whereas the rest have gaps or were started between 2010 and 2015.

\begin{longrotatetable}
\begin{deluxetable}{ccc|ccccccc|c|cc|cc|c|c}
\tablenum{1}
\tablecaption{32 M Dwarfs in the ATLAS Sample}
\centerwidetable
\tabletypesize{\scriptsize}
\label{tab:ATLAS32}
\tablehead{
\colhead{} & \colhead{R.A.} & \colhead{Decl.} & \multicolumn{7}{c}{\textit{RECONS}}  & \colhead{} & \multicolumn{2}{c}{\textit{PDCSAP}} & \multicolumn{2}{c}{\textit{unpopular}} & \colhead{\textit{\# of}} & \colhead{\textit{TESS}}\\[-1em] 
\colhead{Name} & \colhead{J2000.0} & \colhead{J2000.0} & \colhead{\textit{V}} & \colhead{\textit{R}} 
& \colhead{\textit{I}} & \colhead{Filter} & \colhead{Coverage} & \colhead{$\sigma$} & \colhead{IDR} & \colhead{TIC ID} & \colhead{$\sigma_{avg}$} & \colhead{IDR$_{avg}$} & \colhead{$\sigma_{avg}$} & \colhead{IDR$_{avg}$} & \colhead{Sectors} &\colhead{Blending}  \\[-1em] 
\colhead{} & \colhead{} & \colhead{} & \colhead{(mag)} & \colhead{(mag)} 
& \colhead{(mag)} & \colhead{} & \colhead{(yrs)} & \colhead{(mmag)} & \colhead{(mmag)} & \colhead{} & \colhead{(mmag)} & \colhead{(mmag)} & \colhead{(mmag)} & \colhead{(mmag)} & \colhead{} &\colhead{}  \\[-1em] 
\colhead{(1) } & \colhead{(2)} & \colhead{(3)} & \colhead{(4)} & \colhead{(5)}  & \colhead{(6)}  & \colhead{(7)} & \colhead{(8)} &  \colhead{(9)} & \colhead{(10)} & \colhead{(11)} & \colhead{(12)} &
\colhead{(13)}& \colhead{(14)} & \colhead{(15)} & \colhead{(16)} & \colhead{(17)}\\[-2em] 
}
\startdata
GJ 1002 & 00 06 43.2 & $-$07 32 17 & 13.84 & 12.21 & 10.21 & \textit{R} & 19.1 & 11.0 & 25.2 & 176287658 &  &  &  &  & N.O. &  \\
LHS 1140 & 00 44 59.3 & $-$15 16 17 & 14.18 & 12.88 & 11.19 & \textit{V} & 19.0 & 12.7 & 29.1 & 92226327 & 0.9 & 1.7 & 2.2 & 5.6 & 1 & Minor \\
GJ 54.1 & 01 12 30.6 & $-$16 59 56 & 12.16 & 10.72 & 8.94 & \textit{V} & 19.1 & 19.0 & 43.8 & 439403362 & 1.1 & 1.4 & 5.0 & 12.9 & 1 & No \\
GJ 83.1 & 02 00 13.0 & +13 03 07 & 12.35 & 10.95 & 9.18 & \textit{V} & 12.2 & 13.8 & 31.3 & 404715018 &  &  &  &  & N.O.  & \\
TEE0253+1652 & 02 53 00.9 & +16 52 53 & 15.14 & 13.03 & 10.65 & \textit{I} & 19.4 & 5.8 & 14.1 & 257870150 &  &  &  &  & N.O.  & \\
LP 771-95A & 03 01 51.4 & $-$16 35 36 & 11.22 & 10.07 & 8.66 & \textit{V} & 23.2 & 11.1 & 29.3 & 98796344 & 2.4 & 6.5 & 1.2 & 3.1 & 1 & Triple \\
GJ 1057 & 03 13 22.9 & +04 46 29 & 13.94 & 12.45 & 10.62 & \textit{R} & 13.0 & 20.4 & 53.3 & 328465904 & 0.5 & 1.3 & 9.8 & 26.1 & 1 & No \\
GJ 1061 & 03 35 59.7 & $-$44 30 46 & 13.09 & 11.45 & 9.47 & \textit{R} & 23.3 & 22.0 & 59.9 & 79611981 & 0.2 & 0.6 & 0.5 & 1.2 & 2 & No \\
LHS 1678 & 04 32 42.6 & $-$39 47 12 & 12.48 & 11.46 & 10.26 & \textit{V} & 19.0 &  5.4 & 13.2 & 77156829 & 0.4 & 1.0 & 0.7 & 1.8 & 2 & No \\
LHS 1723 & 05 01 57.4 & $-$06 56 46 & 12.20 & 10.86 & 9.18 & \textit{V} & 23.3 & 20.6 & 53.6 & 43605290 & 2.0 & 1.4 & 5.0 & 12.2 & 1 & No \\
LHS 1748 & 05 15 46.7 & $-$31 17 45 & 12.08 & 11.06 & 9.83 & \textit{V} & 22.3 & 12.9 & 27.9 & 77897915 & 0.4 & 1.0 & 2.4 & 6.6 & 2 & Major \\
GJ 273 & 07 27 24.5 & +05 13 33 & 9.88 & 8.68 & 7.14 & \textit{V} & 13.3 & 19.1 & 51.2 & 318686860 & 0.1 & 0.3 & 0.5 & 1.2 & 1 & No \\
L 34-26 & 07 49 12.7 & $-$76 42 07 & 11.31 & 10.19 & 8.79 & \textit{V} & 16.9 & 22.7 & 49.3 & 272232401 & 9.7 & 21.7 & 8.4 & 17.8 & 8 & Minor \\
L 98-59 & 08 18 07.6 & $-$68 18 47 & 11.71 & 10.61 & 9.25 & \textit{R} & 17.1 & 8.1 & 18.5 & 307210830 & 0.4 & 0.7 & 1.2 & 3.0 & 7 & No \\
GJ 317 & 08 40 59.2 & $-$23 27 23 & 12.01 & 10.84 & 9.37 & \textit{R} & 14.3 & 21.0 & 58.5 & 118608254 & 0.7 & 1.4 & 5.9 & 14.9 & 1 & No \\
GJ 357 & 09 36 01.6 & $-$21 39 39 & 10.92 & 9.86 & 8.57 & \textit{V} & 18.3 & 12.4 & 29.1 & 413248763 & 0.2 & 0.5 & 0.5 & 1.0 & 1 & No \\
GJ 367 & 09 44 29.8 & $-$45 46 35 & 10.12 & 9.10 & 7.86 & \textit{V} & 18.3 & 15.4 & 37.6 & 34068865 & 0.4 & 0.9 & 1.8 & 4.6 & 1 & Minor \\
LHS 281 & 10 14 51.8 & $-$47 09 24 & 13.49 & 12.26 & 10.69 & \textit{R} & 22.1 & 13.3 & 34.1 & 101955023 & 0.7 & 1.6 & 1.4 & 3.8 & 2 & Minor \\
LHS 2335 & 10 58 35.1 & $-$31 08 38 & 11.93 & 10.90 & 9.63 & \textit{V} & 22.1 & 11.0 & 26.5 & 49064384 & 0.3 & 0.8 & 1.5 & 4.1 & 1 & No \\
GJ 486 & 12 47 56.6 & +09 45 05 & 11.42 & 10.22 & 8.68 & \textit{V} & 13.1 & 9.8 & 21.5 & 390651552 & 0.4 & 0.8 & 1.4 & 3.6 & 1 & No \\
Proxima Cen & 14 29 42.9 & $-$62 40 46 & 11.13 & 9.45 & 7.41 & \textit{V} & 22.9 & 35.8 & 93.6 & 388857263 & 2.6 & 2.3 & 4.9 & 12.3 & 1 & Minor \\
GJ 581 & 15 19 26.8 & $-$07 43 20 & 10.56 & 9.44 & 8.03 & \textit{V} & 22.9 & 13.2 & 34.9 & 36853511 &  &  &  &  & N.O. & \\
GJ 628 & 16 30 18.1 & $-$12 39 45 & 10.07 & 8.89 & 7.37 & \textit{V} & 19.9 & 11.8 & 31.0 & 413948621 &  &  &  &  & N.O. & \\
GJ 1214 & 17 15 18.9 & +04 57 50 & 14.71 & 13.27 & 11.50 & \textit{I} & 13.1 & 11.1 & 28.3 & 467929202 &  &  &  &  & N.O. & \\
GJ 667C & 17 18 58.8 & $-$34 59 49 & 10.34 & 9.29 & 8.09 & \textit{V} & 20.0 & 9.9 & 22.4 & 154385809 &  &  & 0.7 & 2.1 & 1 & Triple \\
GJ 682 & 17 37 03.7 & $-$44 19 09 & 10.99 & 9.74 & 8.15 & \textit{V} & 20.0 & 13.4 & 27.0 & 16909043 & 0.5 & 1.1 & 2.6 & 6.6 & 1 & Major \\
GJ 1252 & 20 27 42.1 & $-$56 27 25 & 12.20 & 11.19 & 9.93 & \textit{R} & 22.9 & 7.3 & 17.2 & 370133522 & 0.5 & 1.1 & 2.1 & 5.5 & 1 & No \\
GJ 849 & 22 09 40.3 & $-$04 38 27 & 10.38 & 9.27 & 7.87 & \textit{V} & 19.9 & 10.7 & 26.7 & 248027247 &  &  &  &  & N.O. & \\
GJ 1265 & 22 13 42.9 & $-$17 41 09 & 13.63 & 12.31 & 10.60 & \textit{R} & 9.8 & 10.2 & 23.1 & 471012766 &  &  &  &  & N.O. & \\
LHS 3844 & 22 41 58.1 & $-$69 10 08 & 15.26 & 13.74 & 11.88 & \textit{R} & 7.0 & 17.9 & 46.3 & 410153553 & 1.3 & 3.1 & 2.8 & 7.3 & 1 & No \\
GJ 876 & 22 53 16.7 & $-$14 15 49 & 10.18 & 8.97 & 7.40 & \textit{V} & 19.9 & 22.2 & 42.6 & 188580272 & 0.2 & 0.5 & 2.2 & 5.8 & 1 & No \\
2MA2306-0502 & 23 06 29.4 & $-$05 02 29 & 18.79 & 16.52 & 14.10 & \textit{I} & 18.9 & 7.7 & 14.3 & 278892590 &  &  &  &  & N.O. & \\
\enddata

\tablecomments{Columns 1--3 denote the star names and its respective J2000 \textit{Gaia} DR3 coordinates. Columns 4--10 describe aspects of the long-term RECONS data ($\S$4) for each target, while columns 11--17 describe the \textit{TESS} data ($\S$5). N.O. in Column 16 denotes Not Observed. Column 17 gives contamination notes based on blending within the \textit{TESS} pixels.}
\end{deluxetable}
\end{longrotatetable}

\section{\textit{TESS} Mid-term Data AND Results} \label{TESS}

For our study of habitability, it is important to also consider flux changes on planetary surfaces caused by stellar rotation. Rotation occurs on timescales of hours to months, thus it is not easily observable with the RECONS data but can be probed with a higher-cadence observing program. We use \textit{TESS} data to study 23 of the 32 stars in the ATLAS sample that have been observed so far for these mid-term variability signatures.

\subsection{Instrument aboard the TESS spacecraft} \label{TESS Instrument}

\textit{TESS} is an all-sky survey mission launched in 2018 primarily to discover transiting exoplanets around relatively bright and nearby stars \citep{Ricker2015}. The photometric precision of \textit{TESS} scales with the brightness of the target between 60--600 ppm for \textit{TESS} magnitudes of \textit{T} = 6--12 \citep{Stassun2019}. The spacecraft is equipped with four CCD cameras, each of which has a 24 $\times$ 24 degree FOV with pixels that are 21$^{\prime\prime}$ on a side. The filter bandpass covers a broad wavelength range of 600--1000 nm that overlaps most of the Kron-Cousins \textit{R} filter bandpass, and completely encompasses the Kron-Cousins $I$ filter and SDSS $z$ filter (see Figure 1 in \citet{Ricker2015}). It is important to point out that because the \textit{TESS} bandpass is redder than the $V$ and $R$ bands used for the long-term study, variability is generally lower in \textit{TESS} data than in RECONS data because the active regions on M dwarfs are generally hotter than the general photosphere; this is discussed in more detail in $\S$\ref{RECONSvsTESS}.

For \textit{TESS} observations, during its primary mission each hemisphere of the celestial sphere is divided into 13 sectors where each sector spans 6$^{\circ}$ away from the ecliptic up through 12$^{\circ}$ beyond the ecliptic pole. Sectors 1--26 were observed continuously for 27.4 days producing full-frame images (FFIs) at a 30-minute cadence \citep{Sullivan2015}. A sample of 200,000+ targets were pre-selected for faster 2-minute cadence observations in addition to the standard 30-minute observations, and those data were extracted via small image cutouts known as postage stamps or Target Pixel Files (TPFs). For this initial assessment of the planetary environments supplied by the ATLAS sample, we utilize only the primary mission FFIs; we plan to expand the effort to include the extended mission FFIs in future work.

\subsection{Photometric Reductions for TESS Mid-term Data} \label{TESS Method}

Raw FFIs are downloaded and calibrated at the \textit{TESS} Science Processing and Operations Center (SPOC) to remove detector effects. SPOC performs traditional calibration methods such as the removal of bias, dark current, and flat fielding along with pixel-level calibration for correcting cosmic rays \citep{Jenkins2016}. Using the Simple Aperture Photometry (SAP) method, SPOC provides us with the SAP (or raw) flux time series data. To mitigate systematics, the Presearch Data Conditioning (PDC) component of the SPOC pipeline performs several corrections by generating a set of cotrending basis vectors that model the systematic trends present in the ensemble flux data, which is similar to the \textit{Kepler} data reduction algorithm \citep{Stumpe2012,Smith2012}. After removing these trends from the time series data and performing SAP on the processed data, SPOC also provides us with the PDCSAP (or processed) flux time series data. The commonly used \texttt{lightkurve} \citep{Barentsen2021} package extracts light curves from these PDCSAP flux time series data sets, where any long-term stellar variability has been removed.  Thus, typical results from SPOC are {\it not} inherently designed to preserve intrinsic stellar signals. Alternative pipelines like \texttt{eleanor} \citep{Feinstein2019} and the MIT Quick-Look Pipeline (QLP) \citep{Huang2020} can be used to extract light curves from raw FFIs. Still, as with the results from SPOC, both pipelines are optimized to detect planet transits and remove low-frequency astrophysical signals. 

The \texttt{unpopular} \citep{Hattori2022} package is an alternative \textit{TESS} pipeline optimized for detrending non-astrophysical systematics in \textit{TESS} FFIs without removing any intrinsic stellar signals from the light curves. The package can preserve stellar rotation signals while removing systematics by simultaneously fitting a polynomial component to capture these astrophysical variations for slowly rotating stars. Here fast and slow rotators refer to stars with rotation periods shorter and longer than 14 days, respectively, which corresponds to roughly half of the 27.4 days observation period of each \textit{TESS} sector. We find that for fast rotators, not including the polynomial component yields a more accurate light curve (discussed further in $\S$ \ref{TESS White Dwarfs}). When using \texttt{unpopular}, rectangular apertures are drawn on target stars that match as closely as possible to the optimal apertures from the SPOC pipeline. 

\begin{figure}
\epsscale{1.1}
\plotone{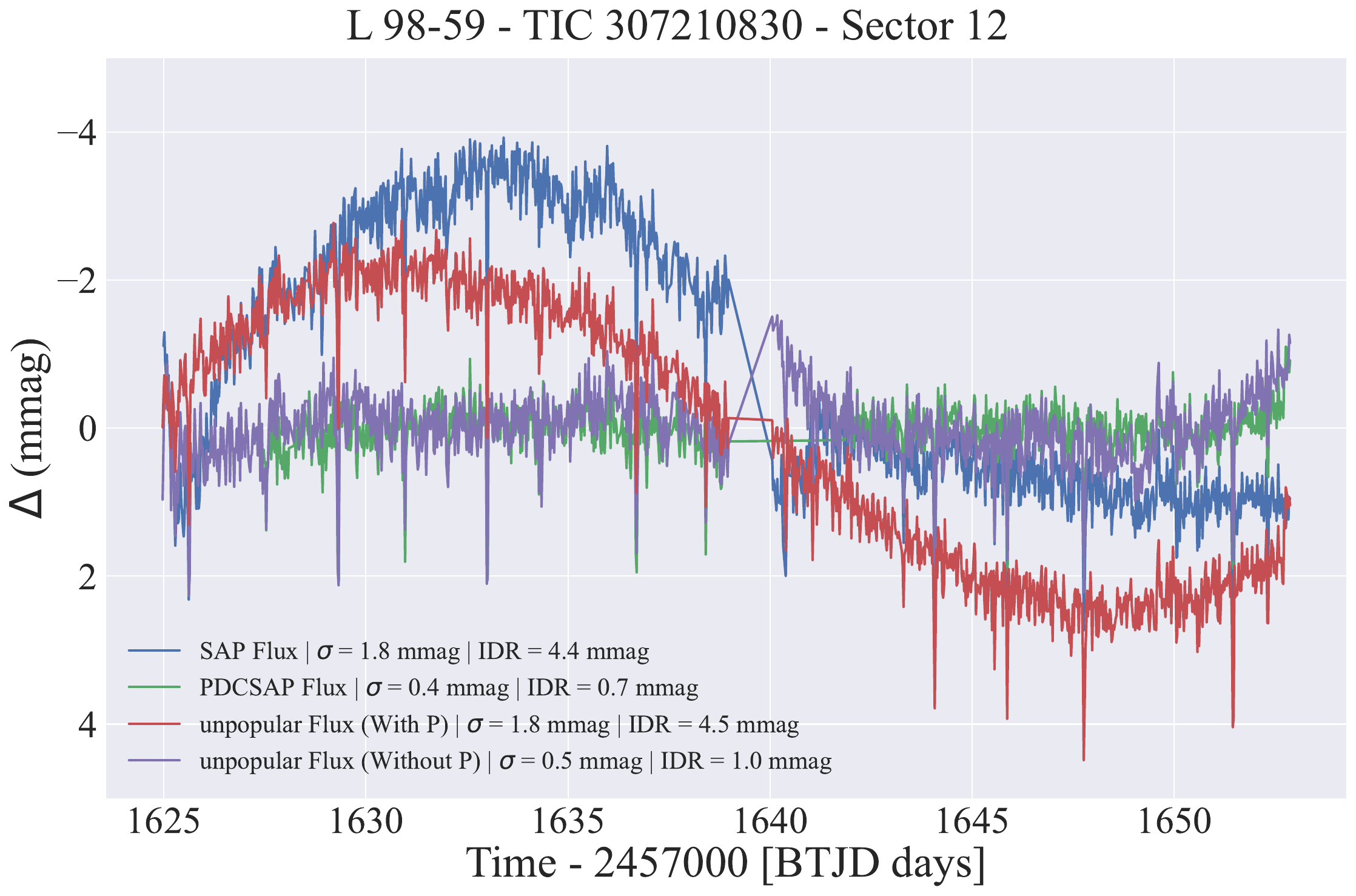}
\caption{L 98-59 \textit{TESS} light curve from sector 12 over 27.4 days. The raw/SAP (blue) fluxes, processed/PDCSAP (green) fluxes, and the \texttt{unpopular} fluxes are overlaid with (red) and without (purple) the inclusion of the polynomial component, respectively. The zero y-axis is defined the same as Figure 3. Note that the rotational modulation of the star is evident in the raw data, eliminated in the PDCSAP reduction, effectively lost in the \texttt{unpopular} reduction without the polynomial, but clear in the \texttt{unpopular} reduction with the polynomial applied. For comparison, sigma and IDR values for variability over the 27 days are given for each reduction in the lower left. P in the legend refers to polynomial.}
\label{fig:L098_059}
\end{figure}

To demonstrate \texttt{unpopular} vs.~the default \textit{TESS} fluxes, we consider L 98-59, an M dwarf with an exoplanet that is also a member of our sample of 32 systems. Figure \ref{fig:L098_059} illustrates the differences between SAP (blue), PDCSAP (green), and \texttt{unpopular} (red and purple) light curves of L 98-59 observed by \textit{TESS} during sector 12. The rotation signal can be seen in the raw SAP light curve (blue) along with the systematic noise, but is removed from the PDCSAP light curve (green) entirely because of the processing techniques. Even with \texttt{unpopular}, the stellar rotation signal is effectively lost when the polynomial component is not included (purple). However, with the inclusion of the polynomial component (red), the rotation signal can be extracted. We find the variability (IDR) for this example to be 4.5 mmag.

\subsection{Testing \texttt{unpopular} with White Dwarfs and M Dwarfs} \label{TESS White Dwarfs}

We carry out two tests to verify the integrity of our methods in applying \texttt{unpopular}, evaluating white dwarfs that exhibit minimal photometric variability to confirm that no systematics are introduced by our techniques, and M dwarfs with measured rotation periods to confirm when to apply, and when not to apply, the polynomial component in \texttt{unpopular}. Details of the targets, observations, and derived IDR values from the various processing results are given in Table \ref{tab:Comparisons}.

\begin{figure}
\epsscale{1.17}
\plottwo{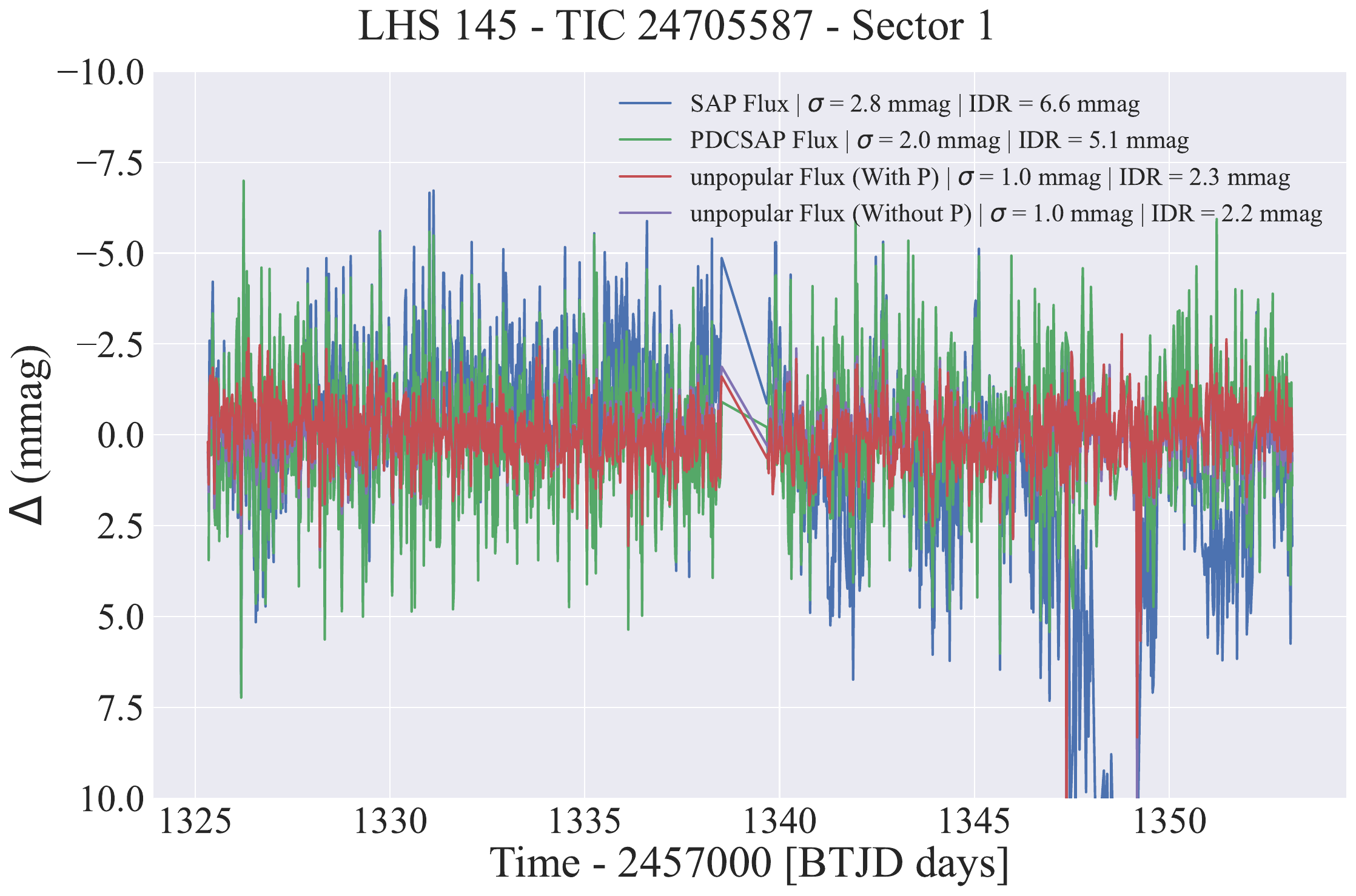}{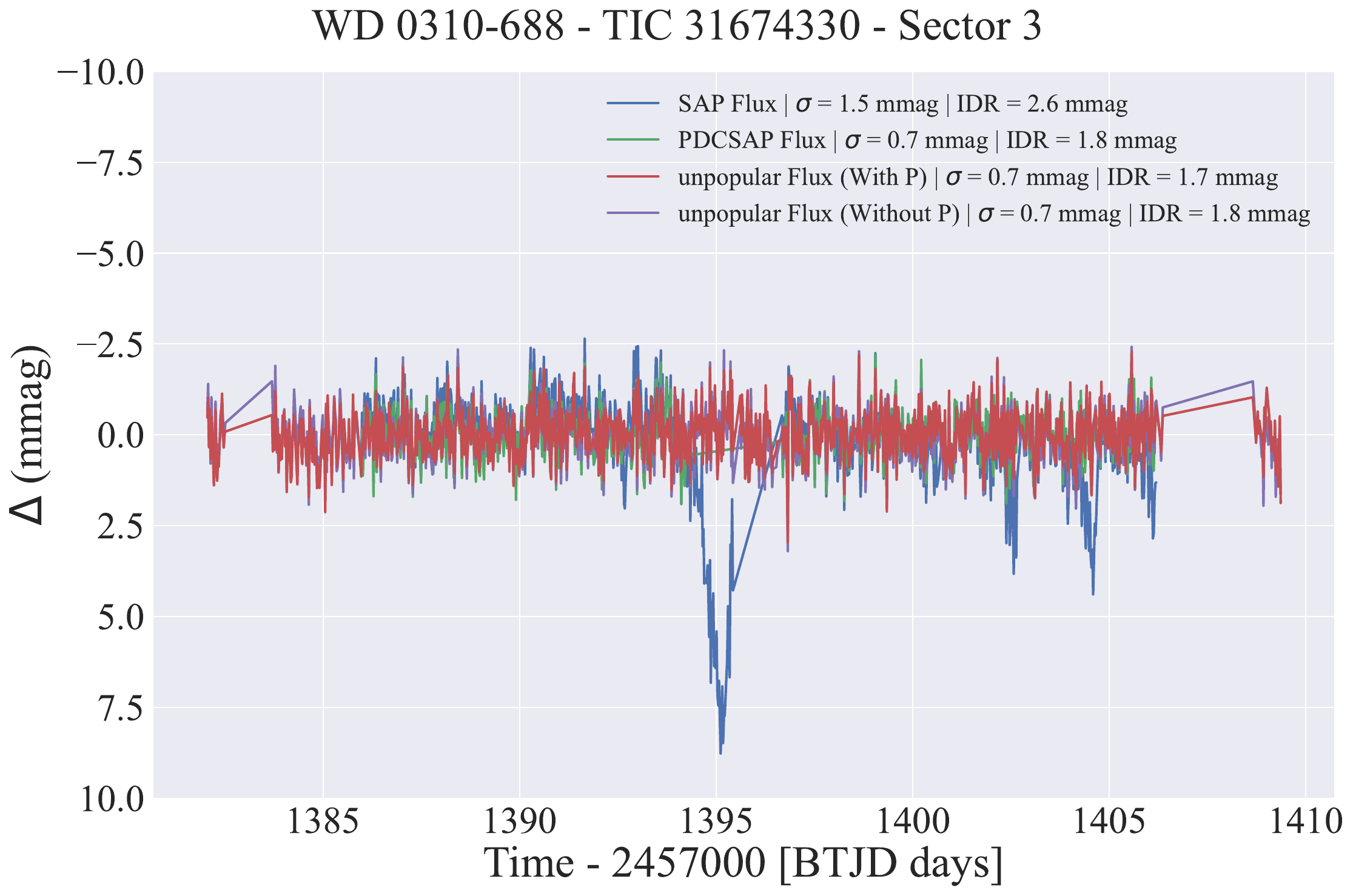}
\plottwo{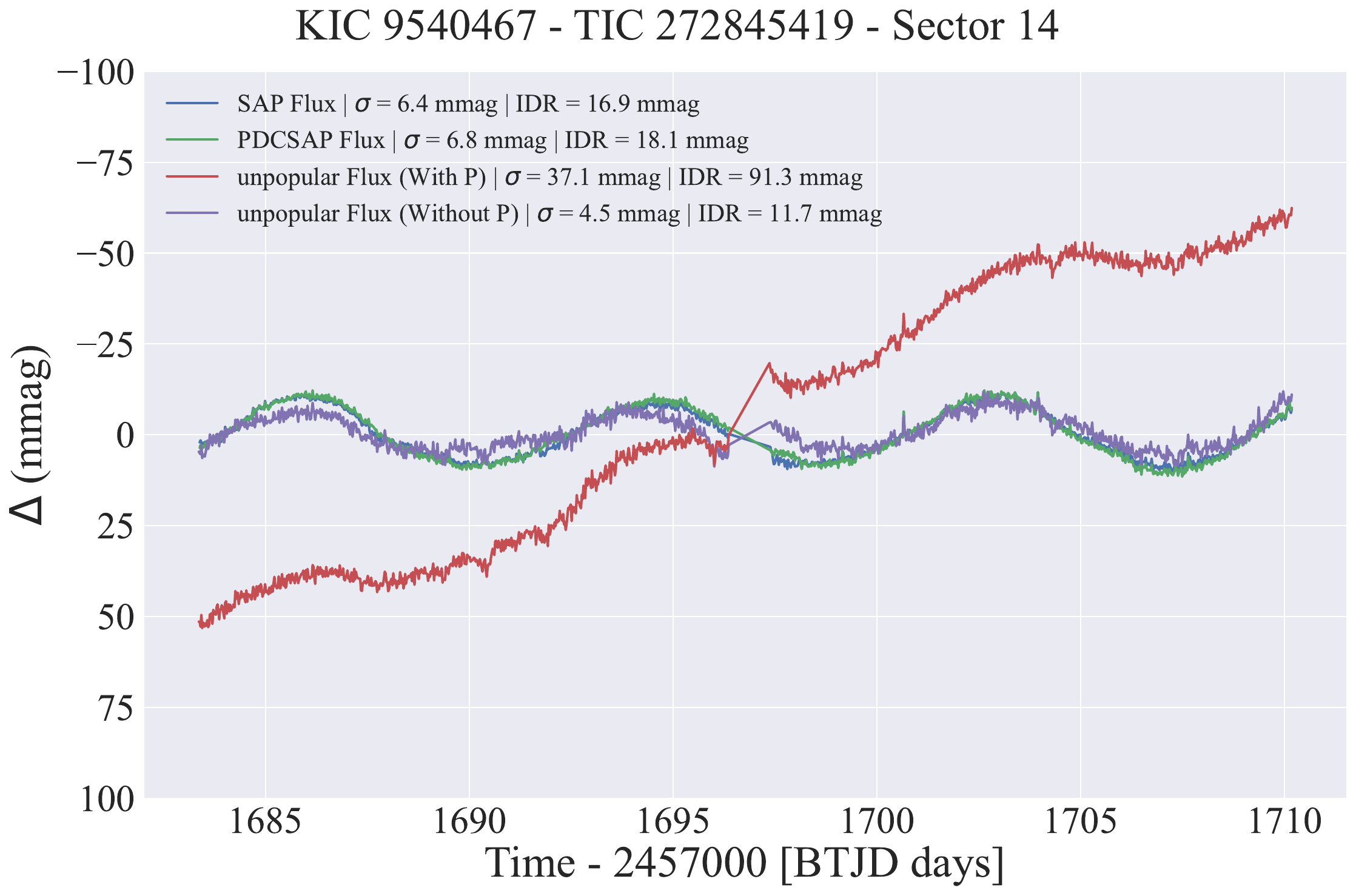}{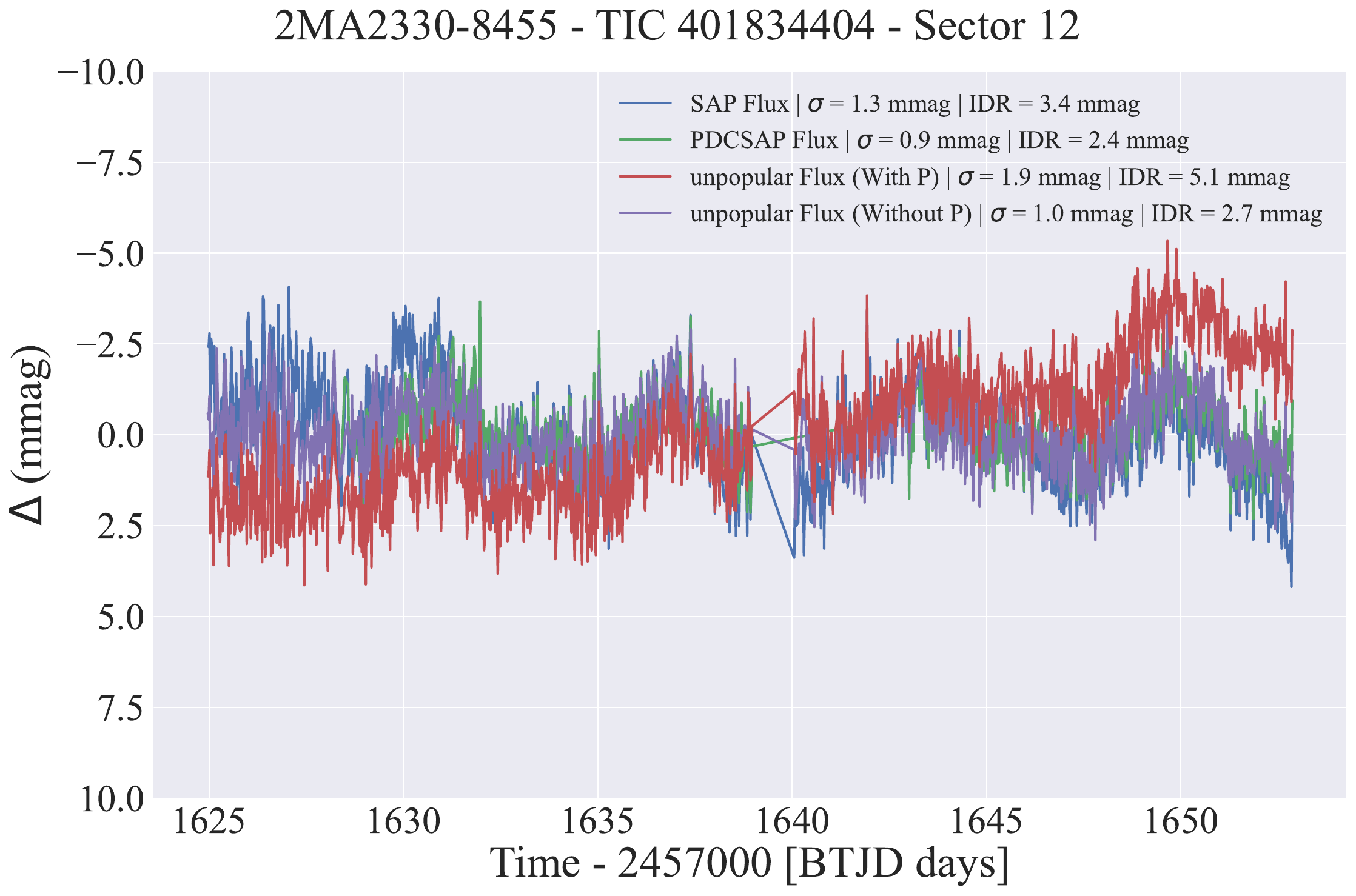}
\plottwo{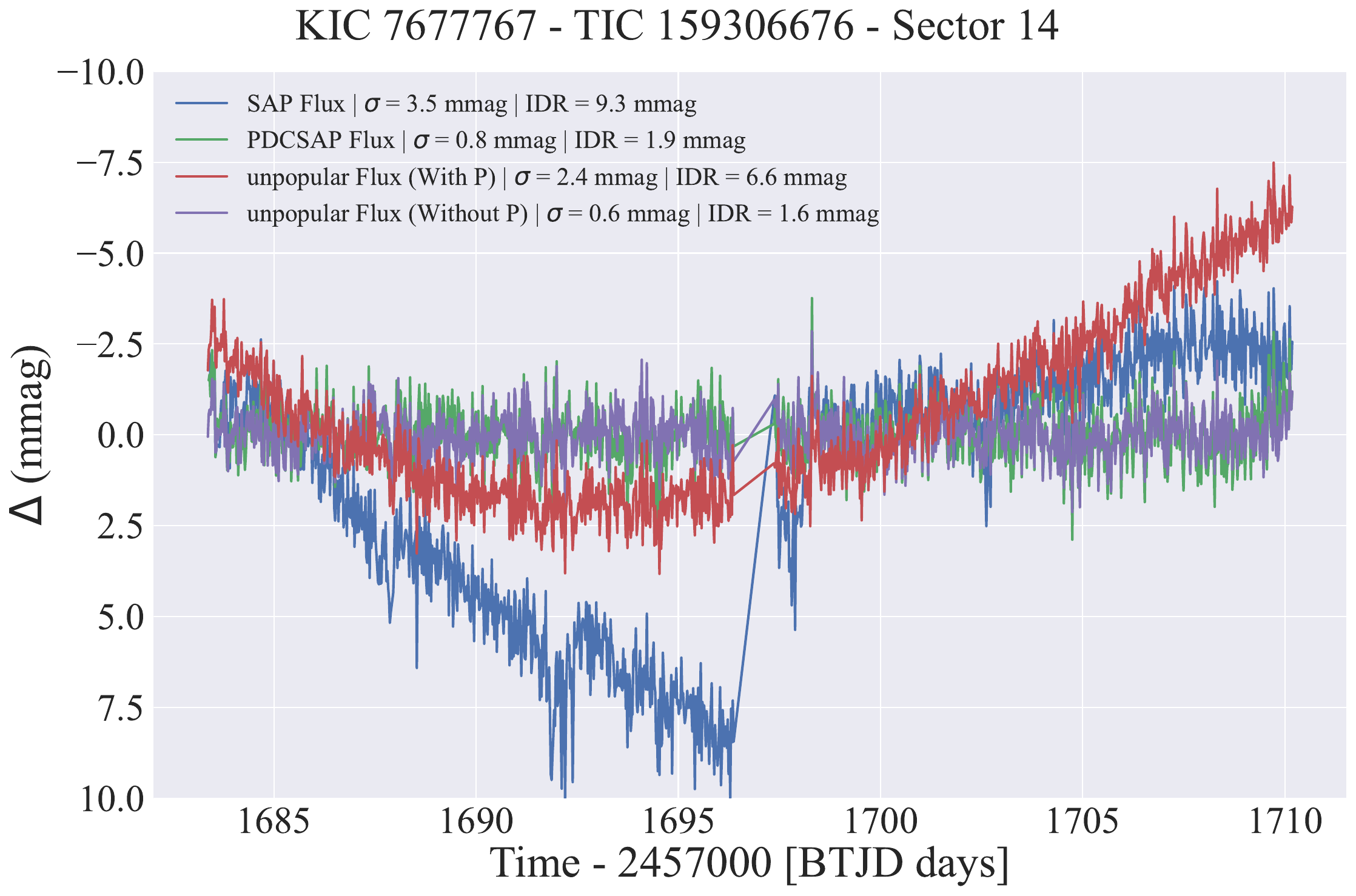}{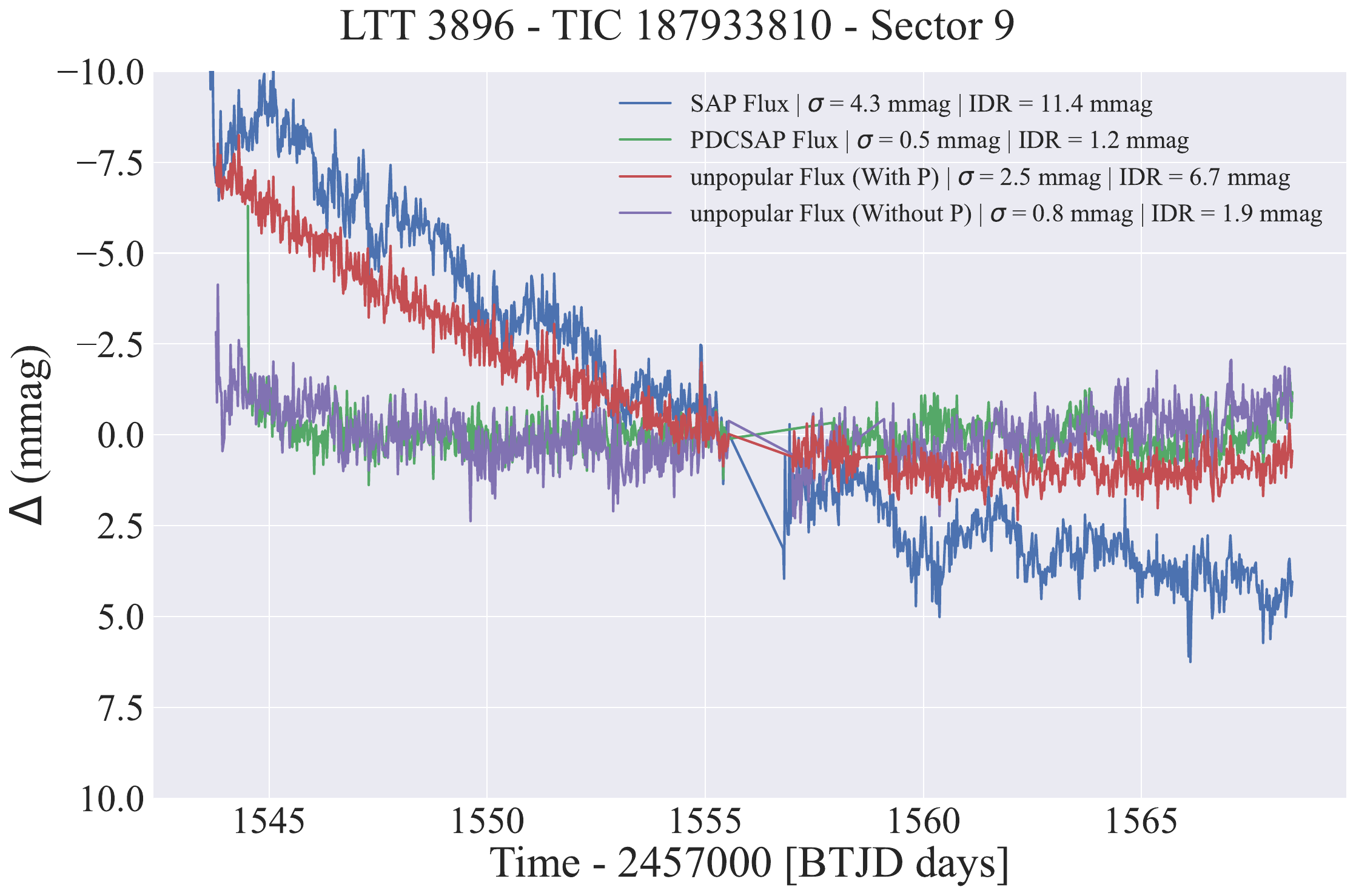}
\caption{
\textbf{top}: \textit{TESS} light curve of LHS 145 (left) and WD 0310-688 (right), two white dwarfs showing no significant photometric variations in fluxes from all four data processing methods (blue, green, red, purple; as defined in Figure \ref{fig:L098_059}). The \texttt{unpopular} without-polynomial line (purple) is similar to the with-polynomial line (red) and therefore is not visible in the plot. No systematics have been introduced in our application of \texttt{unpopular}, with or without the polynomial for non-variable stars.
\textbf{middle}: \textit{TESS} light curves of a fast-rotating 8.4 days \textit{Kepler} target (left) and a fast-rotating MEarth target (right) with a known rotation period of 6.4 days where the exclusion of the polynomial component (purple) preserves the high frequency rotation signal.
\textbf{bottom}: \textit{TESS} light curves of a slow-rotating 28.1 days \textit{Kepler} target (left) and a slow-rotating MEarth (right) with a known rotation period of 91.7 days where the inclusion of the polynomial component (red) preserves the low frequency rotation signal.
}
\label{fig:comparisons}
\end{figure}

SAP fluxes are known to retain systematics due to spacecraft pointing jitter, momentum dumps, focus changes, long-term pointing drifts, etc. These systematics can manifest as flux discontinuities, sudden ramp up / down flux levels, or very short period non-astrophysical flux changes. Some of these can be seen in the SAP fluxes (in blue) in Figure \ref{fig:comparisons}. To ensure that we are not introducing systematics into our variability measurements, we apply our techniques with \texttt{unpopular} to four known bright white dwarfs that are presumed to be photometrically stable (inactive) and exhibiting no detectable variability \citep{Jao2011,Subasavage2017}. For example, the top row of Figure \ref{fig:comparisons} shows the \textit{TESS} light curve of two white dwarfs: LHS 145 (left) and WD 0310-688 (right), where both of the \texttt{unpopular} light curves (red and purple) in each are effectively flat. We also find a flat light curve for the other two white dwarfs: LD 852-007 and WD 1620-391 (not shown here). These light curves clearly demonstrate that \texttt{unpopular} is better than the SAP fluxes where the systematics are much more obvious. As seen in Table \ref{tab:Comparisons}, we observe that the IDRs for these four white dwarfs are $<$ 4 mmag. 

As a proof of concept in applying our techniques with \texttt{unpopular} to variable stars, we test our methodology on a set of 10 bright M dwarfs with known rotation periods. We selected five M dwarfs with KIC names from \textit{Kepler} \citep{McQuillan2013} and five from the MEarth survey \citep{Newton2018} and list them in Table \ref{tab:Comparisons}.  This set includes three fast and seven slow rotators, with periods spanning a range of three days to three months; reported rotation periods are given in the final column of Table \ref{tab:Comparisons}. The middle row of Figure \ref{fig:comparisons} illustrates the light curve for two (KIC 9540467 and 2MA2330-8455) of the three fast rotators while the bottom row illustrates two (KIC 7677767 and LTT 3896) of the seven slow rotators, showing the SAP, PDCSAP, and \texttt{unpopular} (with and without the polynomial component) fluxes from \textit{TESS}. 

We find that for the three fast rotators: LHS 2836, 2MA2330-8455, and KIC 9540467 including the polynomial (red curve) introduces a false long-term trend that is not seen in the SAP fluxes (blue curve, nearly identical to the PDCSAP curve in green), but the \texttt{unpopular} reduction without the polynomial (purple curve) preserves the true stellar variability as shown in the two examples in the middle row of Figure \ref{fig:comparisons}. We further analyze the chosen \texttt{unpopular} fluxes for these three fast rotators by computing a Lomb-Scargle periodogram. We find that the resulting rotation periods: 3.3, 6.3, and 8.5 days align closely with the reported periods given in Table \ref{tab:Comparisons}. For the seven slow rotators: KIC 7692454, KIC 7677767, KIC 4043389, GJ 1088, KIC 10647081, L 154-205, and LTT 3896, the inclusion of a polynomial component (red curve) smoothed and preserved the long-term signal seen in the SAP fluxes (blue curve), as can be seen in the two examples in the bottom row of Figure \ref{fig:comparisons}.

For all 10 M dwarfs with rotation periods, all four types of reductions were visually inspected to determine when to include, or not include, the polynomial while applying \texttt{unpopular}. It became clear that in cases where we can visually identify a rotation period shorter than half ($\sim$14 days) of the \textit{TESS} observing period in the raw SAP light curves, the polynomial should not be included. For stars without evident rotation shorter than two weeks, the polynomial should be included. For our ATLAS stars, once the decision about the polynomial inclusion has been made, we then determine the photometric variability by measuring the IDR of the respective \texttt{unpopular} fluxes.

\begin{deluxetable*}{cccc|cccc|c}
\tablenum{2}
\tablecaption{Results of Methodology Tests on White Dwarfs and Rotating M Dwarfs}
\centerwidetable 
\tabletypesize{\scriptsize}
\label{tab:Comparisons}
\tablehead{
\colhead{} & \colhead{} & \colhead{} & \colhead{} & \multicolumn{4}{c}{IDR (in mmag)} & \colhead{(days)} \\[-1em] 
\colhead{Name} & \colhead{TIC ID} & \colhead{TESS Mag\tablenotemark{a}} & \colhead{Sector} & \colhead{SAP Flux} & \colhead{PDCSAP Flux} & \colhead{\textit{unpopular} with P} & \colhead{\textit{unpopular} no P} & \colhead{P$_{rot}$\tablenotemark{b}} \\[-1em]
}

\startdata
LHS 145 & 24705587 &13.37 & 1 & 6.6 & 5.1 & 2.3 & 2.2 & --- \\
WD 0310-688 & 31674330 & 11.58 & 3 & 2.6 & 1.8 & 1.7 & 1.8 & --- \\
LP 852-007 & 398243520 & 12.61 & 10 & 6.7 & 3.3 & 3.6 & 3.5 & --- \\
WD 1620-391 & 4400550 & 11.28 & 12 & 5.2 & 1.9 & 2.2 & 1.7 & --- \\
\hline
KIC 9540467 & 272845419 & 11.01 & 14 & 16.9 & 18.1 & 91.3 & 11.7 & 8.4 \\
KIC 7692454 & 271432402 & 11.54 & 14 & 11.4 & 3.5 & 5.2 & 3.7 & 16.5 \\
KIC 7677767 & 159306676 & 11.33 & 14 & 9.3 & 1.9 & 6.6 & 1.6 & 28.1 \\
KIC 4043389 & 121214976 & 10.26 & 14 & 7.1 & 1.2 & 5.5 & 2.2 & 38.9 \\
KIC 10647081 & 48189085 & 10.20 & 14 & 4.2 & 1.1 & 3.7 & 2.2 & 69.7 \\
\hline
LHS 2836 & 125421413 & 10.05 & 11 & 18.5 & 8.4 & 11.8 & 8.3 & 3.3 \\
2MA2330-8455 & 401834404 & 11.53 & 12 & 3.4 & 2.4 & 5.1 & 2.7 & 6.4 \\
GJ 1088 & 231917352 & 9.74 & 5 & 2.6 & 0.8 & 4.3 & 1.0 & 53.7 \\
L 154-205 & 447382925 & 11.32 & 12 & 6.7 & 2.4 & 8.0 & 1.8 & 73.1 \\
LTT 3896 & 187933810 & 10.33 & 9 & 11.4 & 1.2 & 6.7 & 1.9 & 91.7 \\
\enddata

\tablenotetext{a}{\textit{TESS} magnitude from the TESS Input Catalog (TIC) v8.2 \citep{Paegert2021}}
\tablenotetext{b}{Rotation period from \citet{McQuillan2013} (Kepler) and \citet{Newton2018} (MEarth)}

\end{deluxetable*}

\vfil

\subsection{Checks for Contamination in TESS Mid-term Data} \label{TESS Contamination}

\textit{TESS} has a very large pixel scale (21$^{\prime\prime}$ pixel$^{-1}$), so drawing just a 3$\times$3 pixel grid results in a $\sim$1$^{\prime}$ aperture. Nearby companions or background sources that are within these apertures will blend with targeted stars and contaminate their integrated fluxes. Therefore, we use the python package \texttt{tpfplotter} \citep{Aller2020} to check the ATLAS stars observed by \textit{TESS} for any contamination within the rectangular apertures used by \texttt{unpopular}. This tool allows us to overlay our rectangular apertures and \textit{Gaia} DR3 sources onto the TPFs to identify contaminants. We categorize blending into the following three types: 
\begin{enumerate}
    \item Major Blending: Targets that have contaminants with $\Delta G$ $\lesssim$ 2 mag in the chosen aperture.
    \item Minor Blending: Targets that have contaminants with $\Delta G$ between 2 and 4 mag in the chosen aperture.
    \item No Blending: Targets that have no contaminants with $\Delta G$ $<$ 4 mag in the chosen aperture.
\end{enumerate}
 
\begin{figure*}
\gridline{\fig{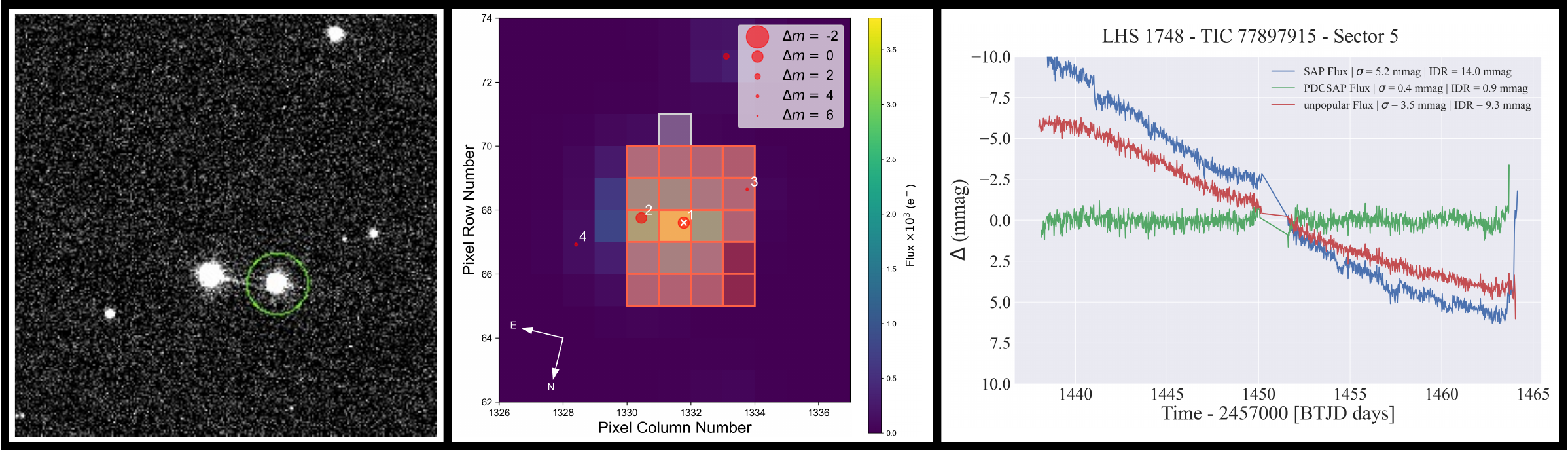}{\textwidth}{(a) LHS 1748: Major Blending}}
\gridline{\fig{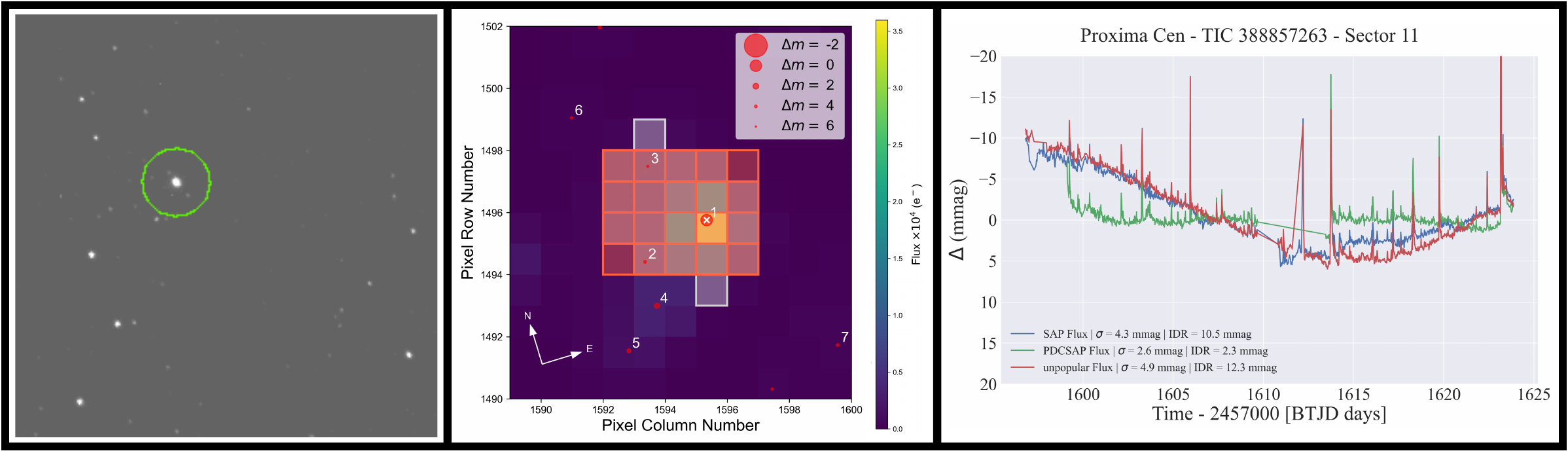}{\textwidth}{(b) Proxima Cen: Minor Blending}}
\gridline{\fig{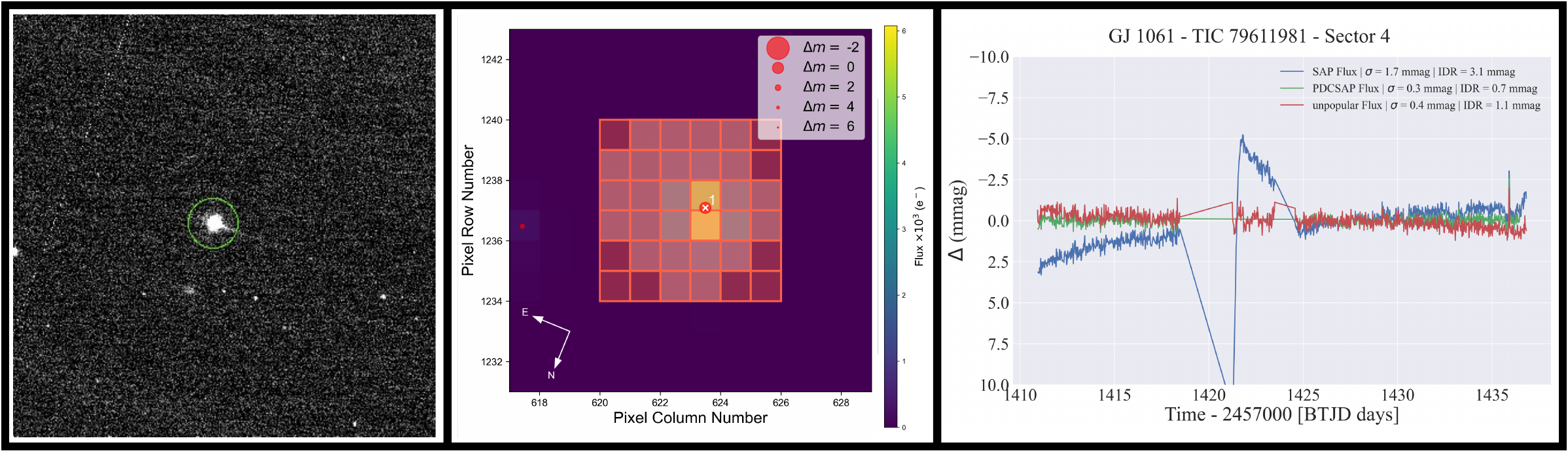}{\textwidth}{(c) GJ 1061: No Blending}}
\caption{Panels illustrating three different types of blending in \textit{TESS}, categorized as major (LHS 1748, top row), minor (Proxima Cen, middle row), and no (GJ 1061, bottom row) blending. \textbf{left column:} Images from the CTIO/SMARTS 0.9 m that are $\sim$3\arcmin~on each side (0\farcs401 pixel$^{-1}$ plate scale) in which the ATLAS stars are circled in green. \textbf{middle column:} Target Pixel Files of the ATLAS stars (indicated with a white X and labeled 1 in each panel) from \textit{TESS}. We overlay two apertures for the given target. The white aperture comes from the default SPOC pipeline. Shown in red is our custom rectangular aperture, drawn for the \texttt{unpopular} package to resemble the white SPOC aperture closely, as can be seen by the over plots of these two semi-transparent apertures. With \texttt{tpfplotter}, we identify contaminants in the red aperture. The filled red circles designate \textit{Gaia} DR3 sources in the field where the radius of the circle is scaled to the $\Delta G$ ($\Delta$m scale shown in each panel) value of the source itself. \textbf{right column:} The \textit{TESS} light curves of the ATLAS stars extracted in the same way as in Figure \ref{fig:L098_059}}
\label{fig:blending}
\end{figure*}

$\Delta G$ above refers to the difference in the \textit{Gaia G} magnitude between that target and the contaminant. Of the 32 ATLAS targets, 23 were identified in \textit{TESS} data and the remaining nine stars are located along the ecliptic, which \textit{TESS} did not observe during its primary mission. Two stars (LHS 1748 and GJ 682) are considered to have major blending, five stars (LHS 1140, L 34-26, GJ 367, LHS 281, and Proxima Cen) have minor blending, while 14 of 23 stars in the sample have no blending. The final two cases, GJ 667C and LP 771-95A, are triple systems unresolved in \textit{TESS} with no additional blending beyond the components named in the systems. Still, given that the IDR values are meant to be considered for the planet host only, we consider those to be major blends because variability could be occurring on any or all of the component stars. Identifying ways to extract only the variability of the exoplanet host from these systems is a subject of our future work. The results of our contamination checks for the 23 ATLAS stars observed by \textit{TESS} are included in Table \ref{tab:ATLAS32}. Examples of each type of blending are shown in Figure \ref{fig:blending}. The top panel shows the major blending case of LHS 1748 because there is a bright contaminant $\sim$0\farcm5 away that is 0.1 mag fainter in $G$. The middle panel shows the minor blending case of Proxima Cen because there are two sources that are $\sim$0\farcm4 and $\sim$0\farcm5 away, but which are 3.9 mag and 4.8 mag fainter in $G$, respectively. The lower panel shows the unblended case of GJ 1061, which has no comparably bright contaminants inside the aperture.

\vfil

\subsection{Results from the TESS Mid-term Data} \label{TESS Results}

\begin{figure}
\epsscale{1.17}
\plottwo{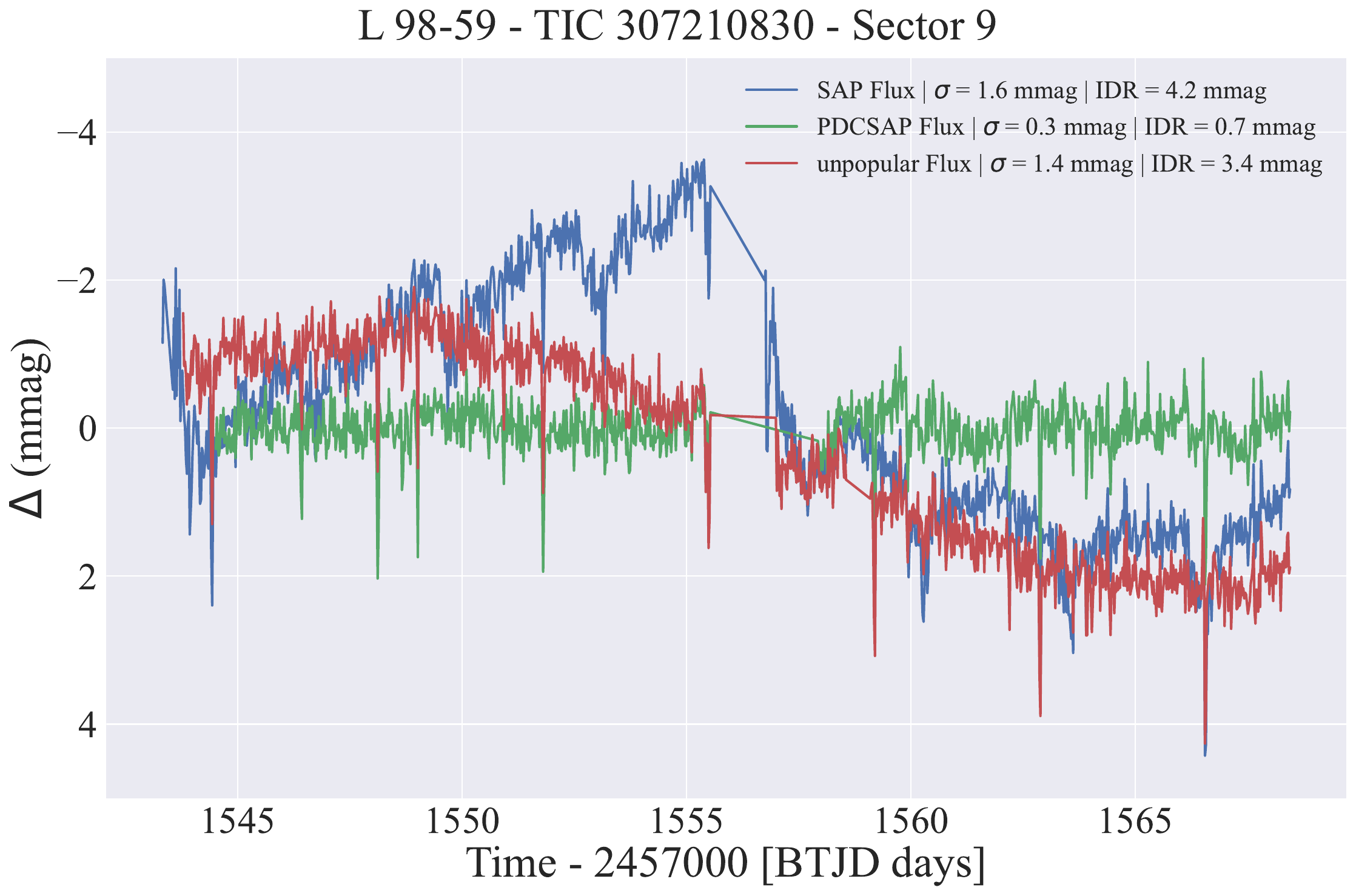}{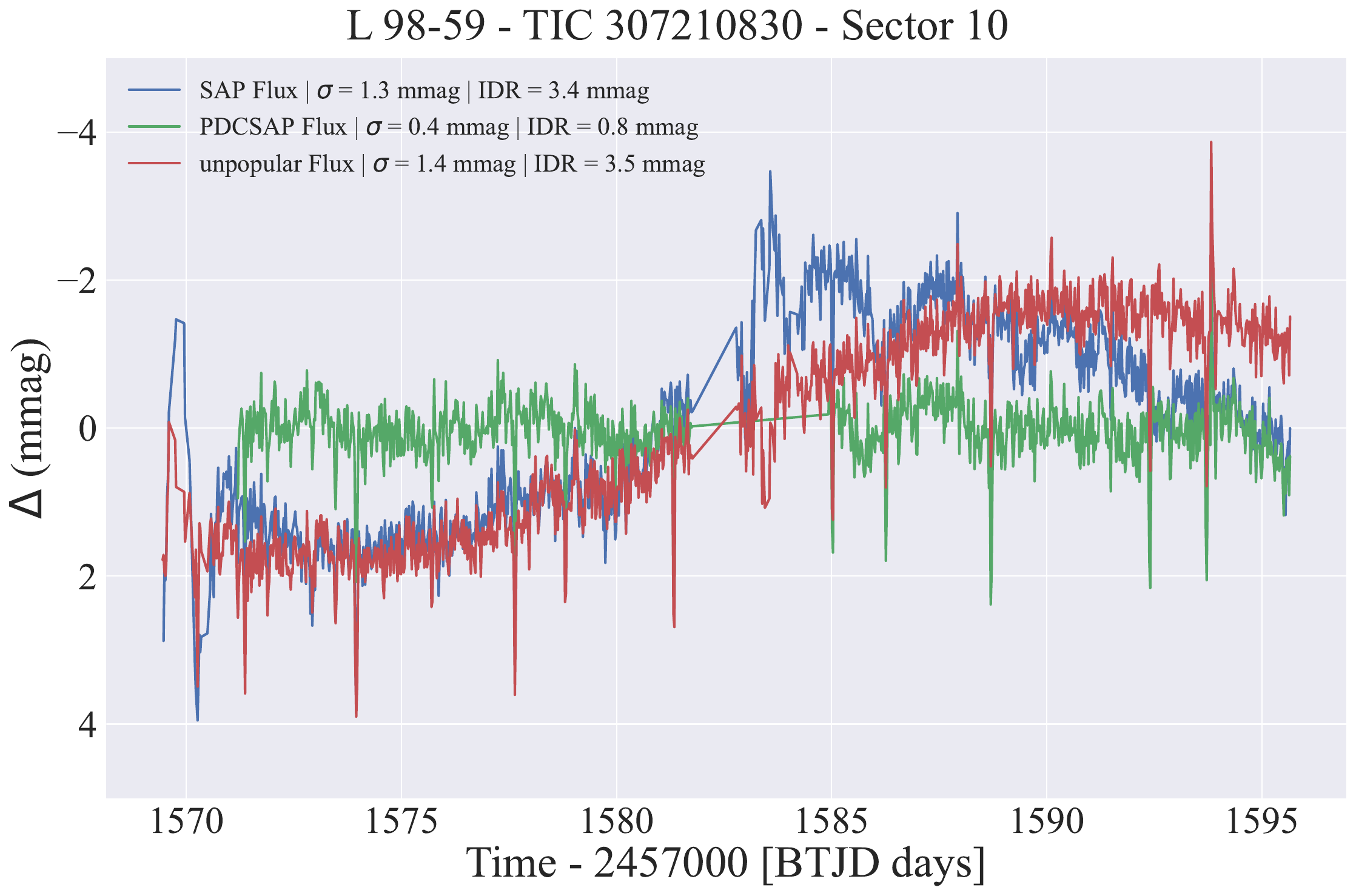}
\plottwo{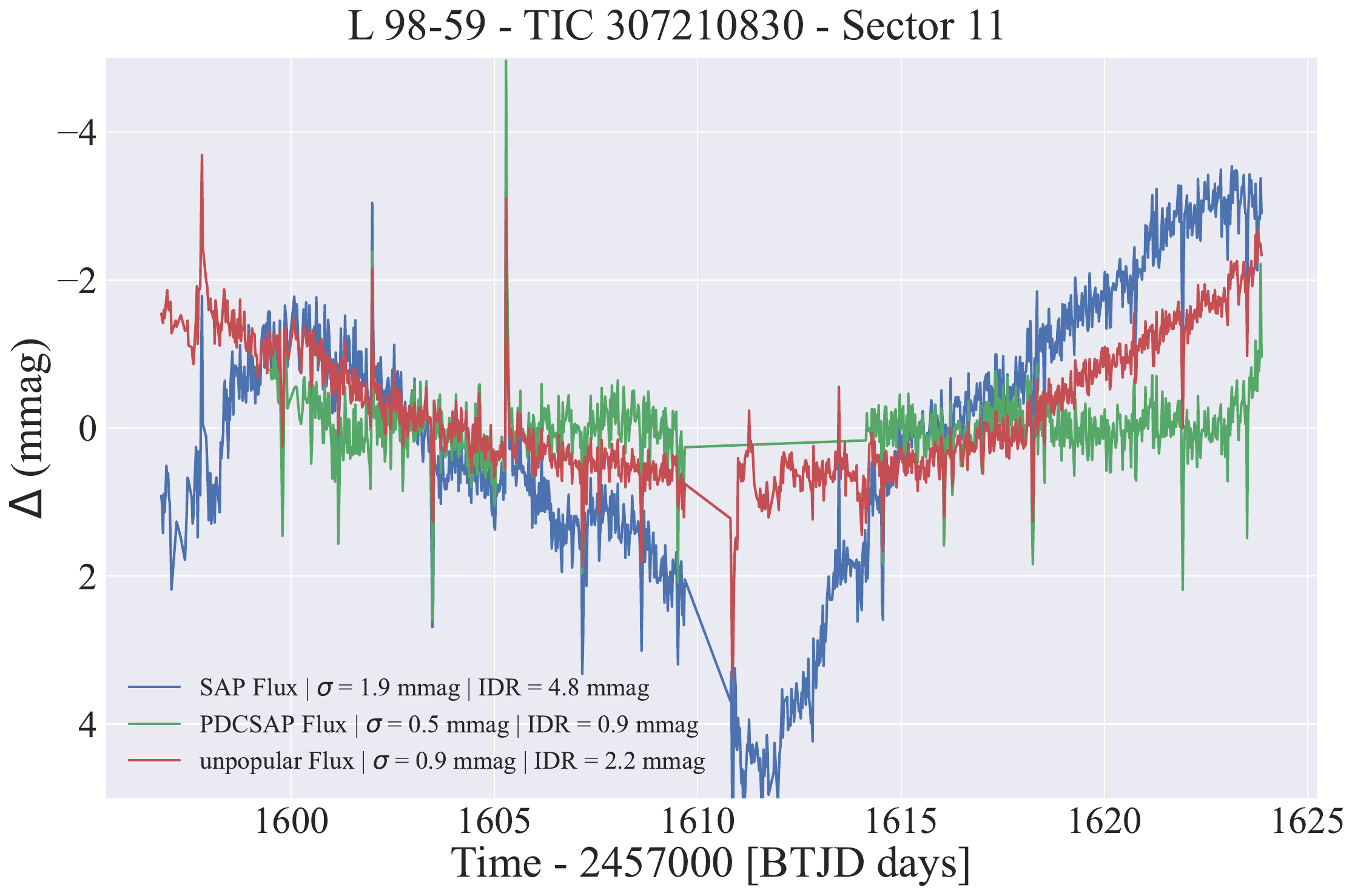}{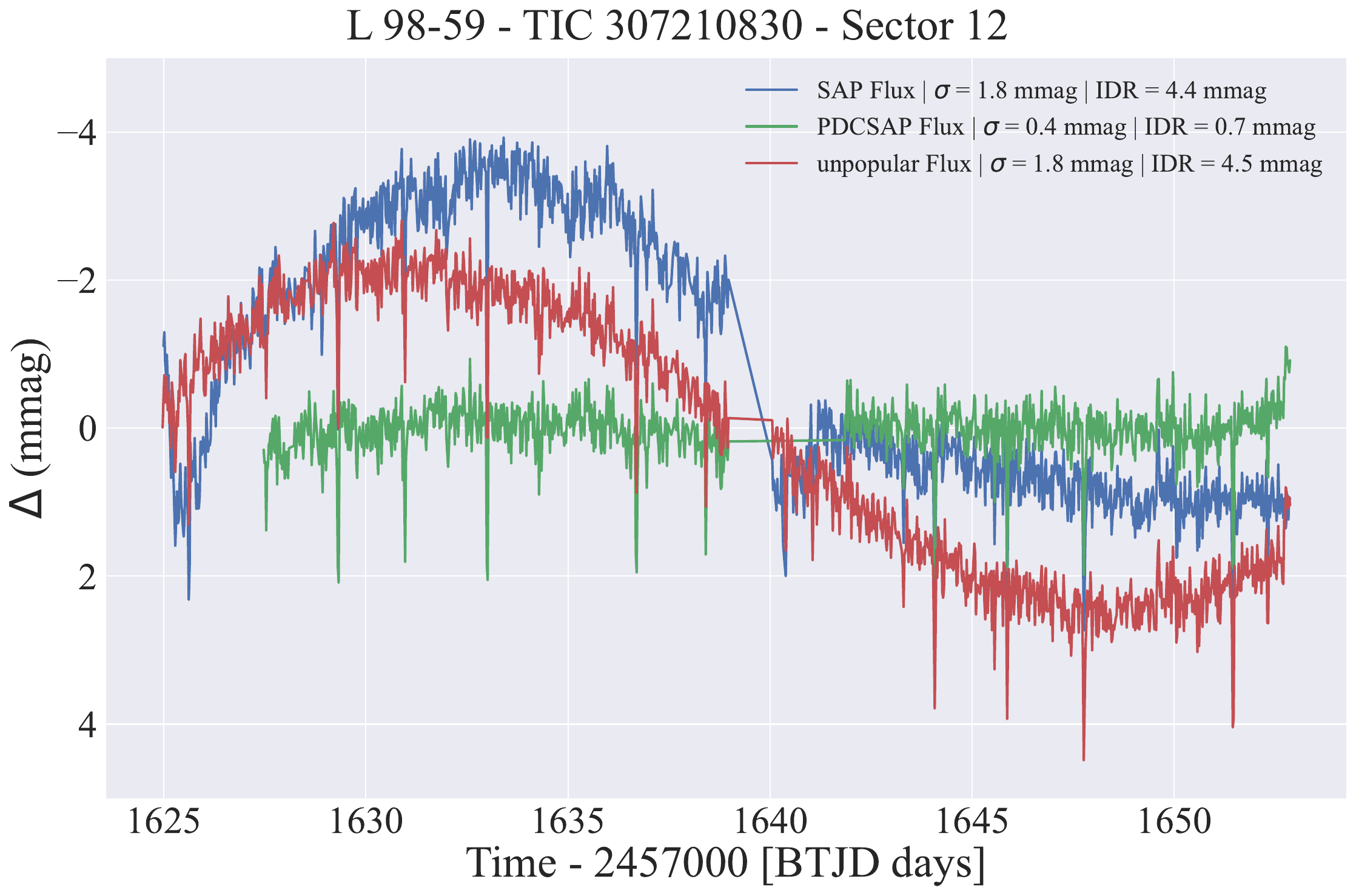}
\caption{\textit{TESS} light curves from selected four consecutive sectors for L 98-59, an intermediate variable in the ATLAS sample. Light curves using the three reductions (blue, green, and red) are as defined in Figure \ref{fig:L098_059}. The IDR value for each sector is given in the legend for the three reductions, but these sectors are not stitched together. Instead, we report the average IDR value in Table \ref{tab:ATLAS32} when multiple sectors like these are available.
}
\label{fig:L098_059_ALL}
\end{figure}

The variability results from \textit{TESS} data are given in the last seven columns (11--17) in Table \ref{tab:ATLAS32}. The TIC ID (11) is followed by four quantities describing the variability results (12--15), where we provide both the $\sigma$ and IDR values from PDCSAP and \texttt{unpopular} reductions so that these quantities may be compared. The number of sectors (16) that cover each star is also noted. Multi-sector stitching is currently difficult and beyond the scope of this work. Typical techniques to stitch sectors, as demonstrated by the often-used \texttt{lightkurve} package, is to normalize the fluxes for each sector and combine all sectors; while this is useful for transit searches, it does not work for astrophysical signals when offsets occur between sectors. Thus, we report the variability directly for single-sector observations, and for multi-sector observations like those for L 98-59 shown in Figure \ref{fig:L098_059_ALL}, we calculate the average of the IDR from all available sectors as its variability. Column 16 of Table \ref{tab:ATLAS32} gives the number of sectors used in the IDR measurements (or N.O.~for no observations) and column 17 notes any type of blending for the targeted star.

\begin{figure}
\epsscale{1.17}
\plottwo{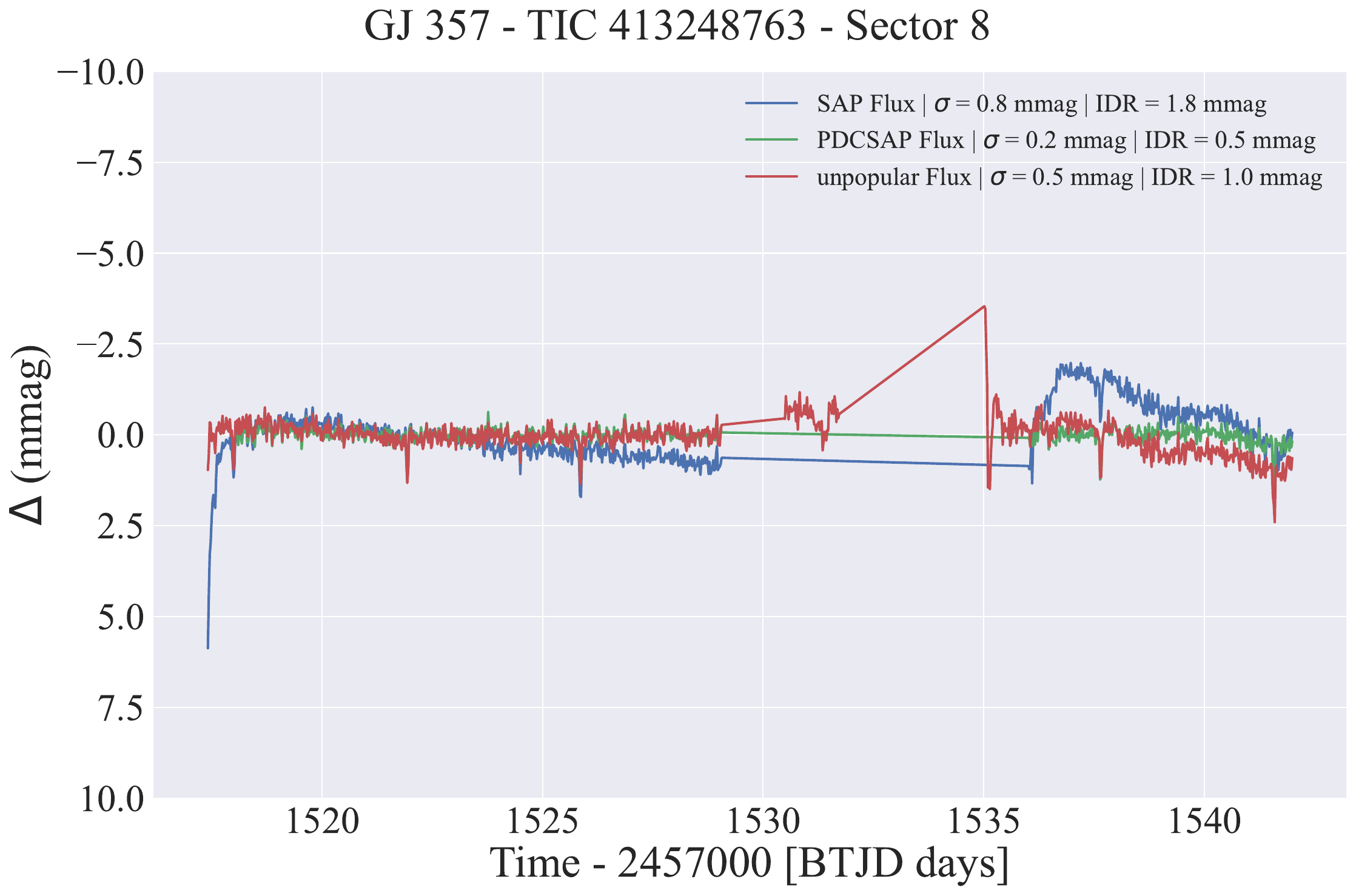}{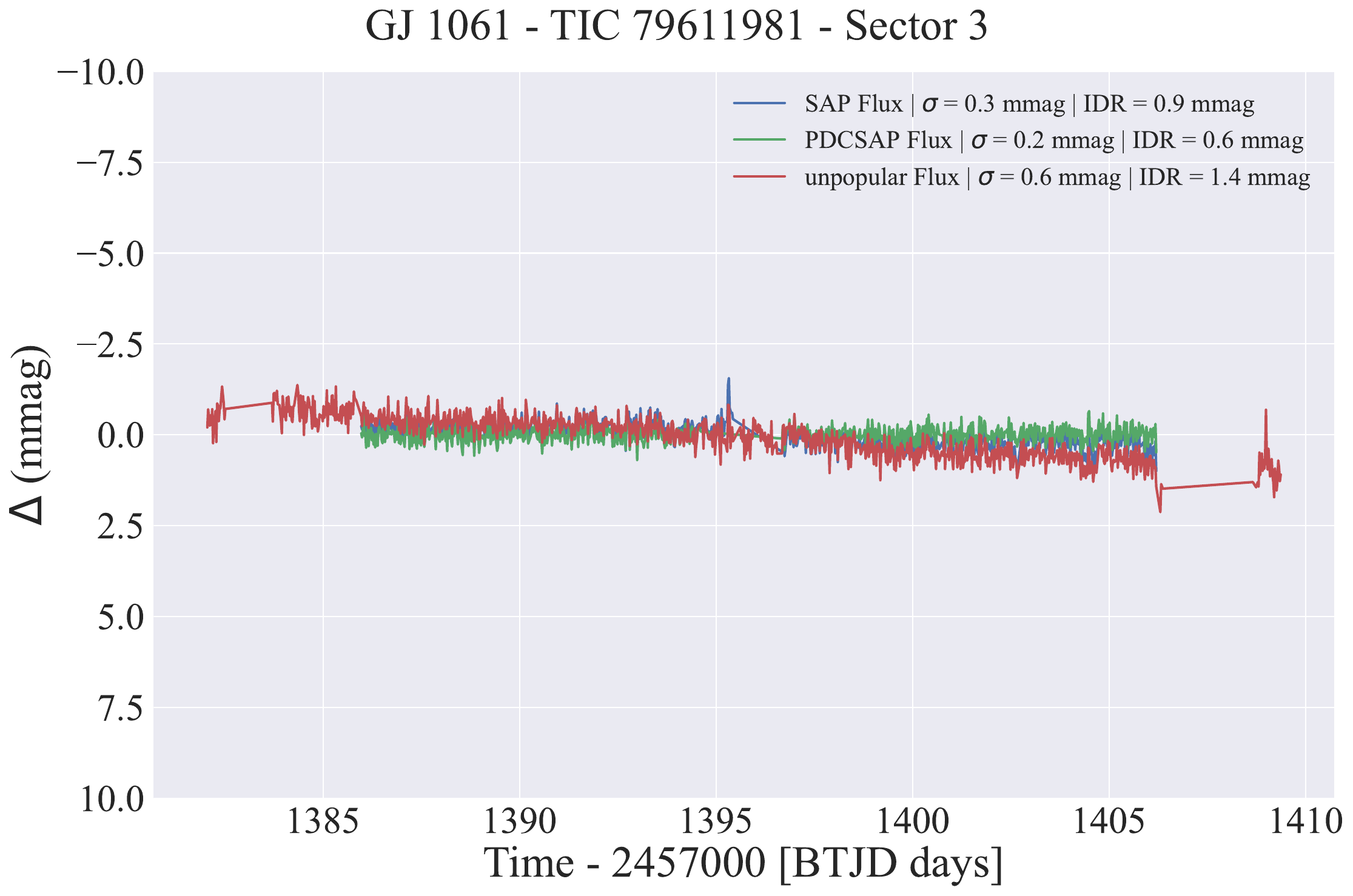}
\caption{\textit{TESS} light curves for the two least variable stars in the ATLAS sample: GJ 357 (left) and GJ 1061 (right). Light curves are shown from reductions with colors as defined in Figure \ref{fig:L098_059}.}
\label{fig:least_TESS}
\end{figure}

\begin{figure}
\epsscale{1.17}
\plottwo{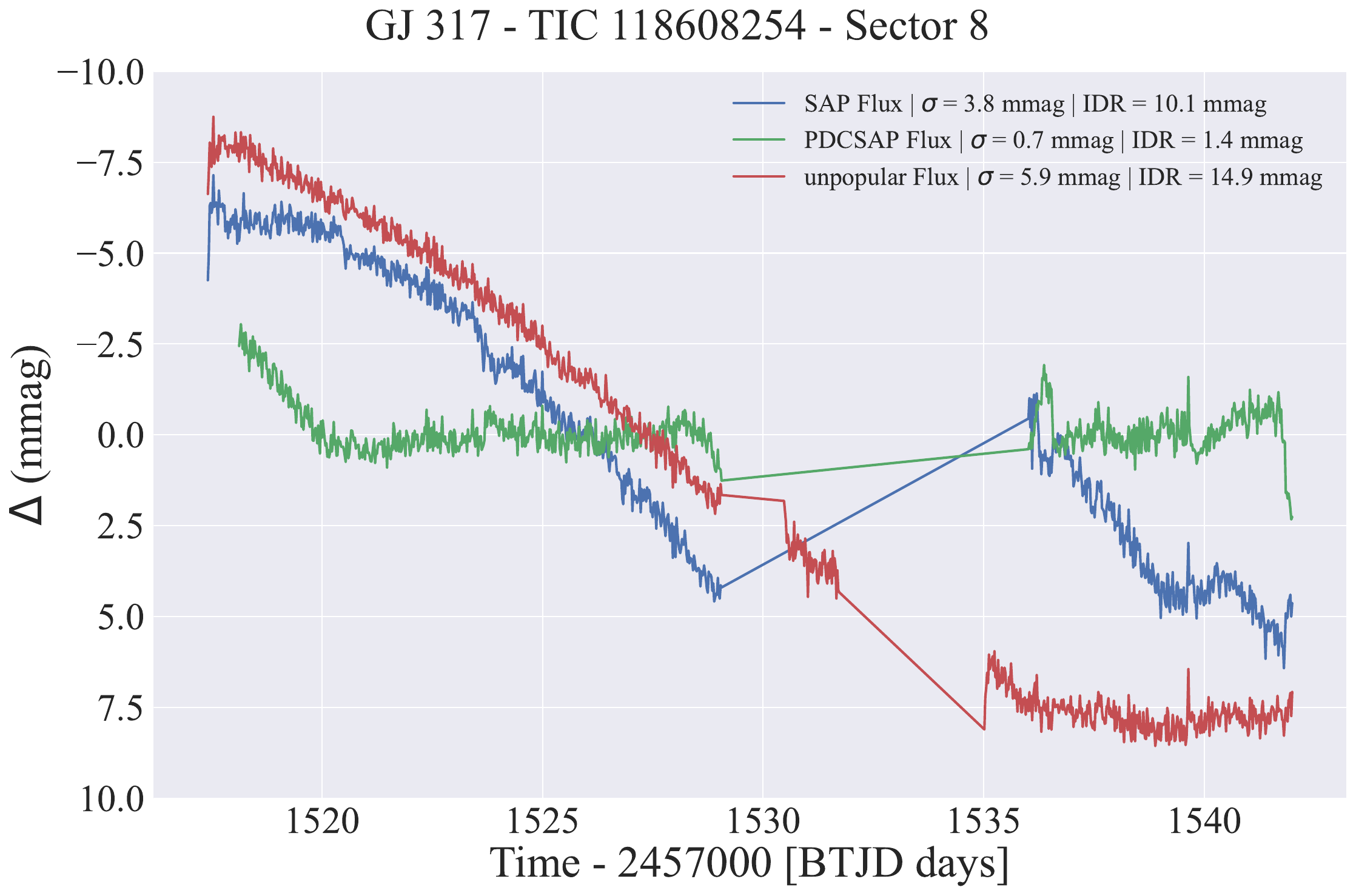}{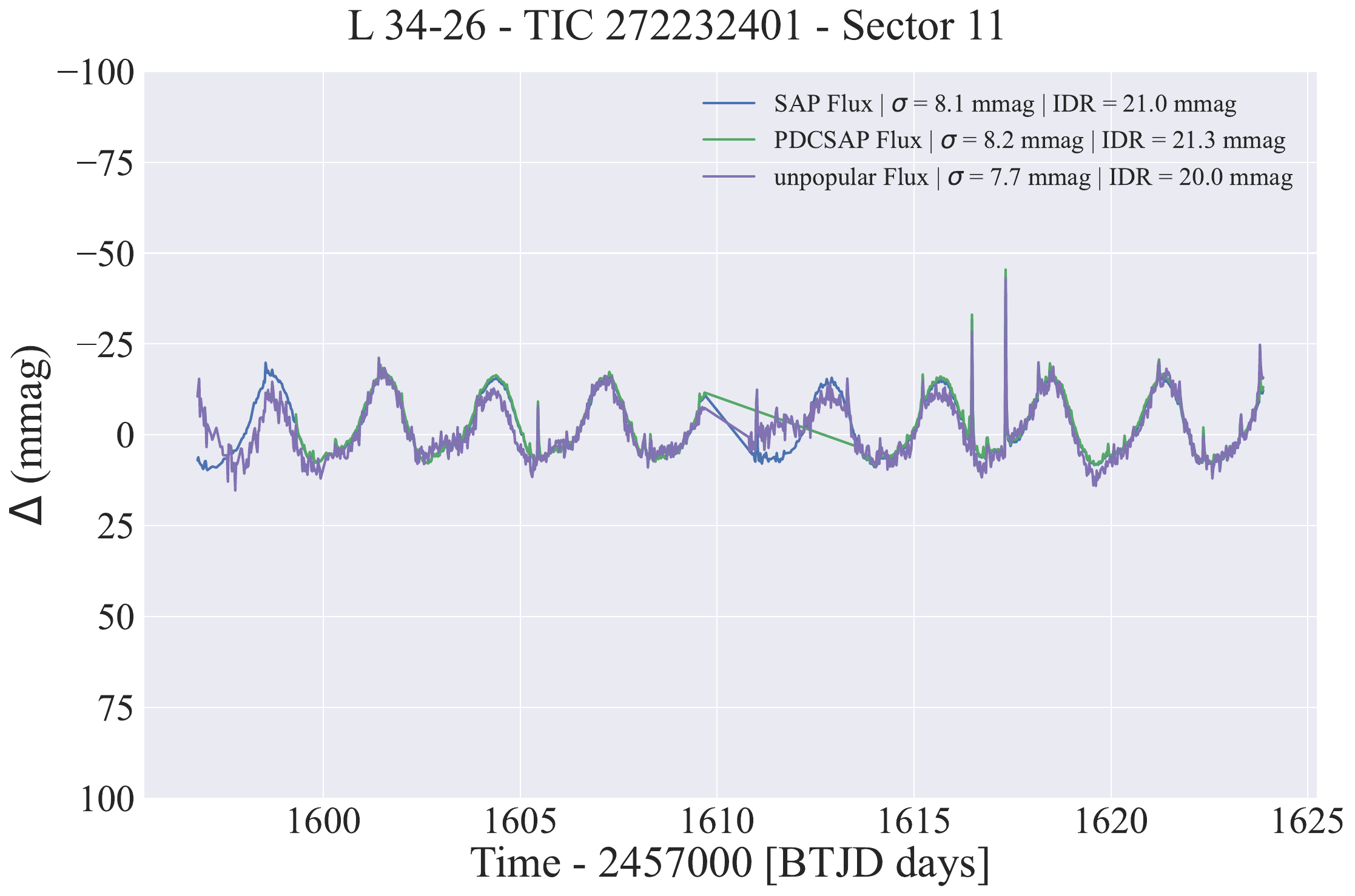}
\caption{\textit{TESS} light curves for the two most variable stars in the ATLAS sample: GJ 317 (left) and L 34-26 (right). Light curves are shown from reductions with colors as defined in Figure \ref{fig:L098_059}. Note the different y-scale for L 34-26 and the purple color of the \texttt{unpopular} flux which is a result of excluding the polynomial component due to the obvious fast rotation signal that is present in the data.}
\label{fig:most_TESS}
\end{figure}

We highlight four stars in Figures \ref{fig:least_TESS} and \ref{fig:most_TESS}, including light curves for two of the least variable stars in Figure \ref{fig:least_TESS} and the two most variable stars in  Figure \ref{fig:most_TESS}.  In all but the fast-rotation case of L 34-26, the PDCSAP light curves (green) tend to be flat because of the removal of the astrophysical signals detrended by the PDC module of the SPOC pipeline, while the \texttt{unpopular} light curves retain those signals. For confirmation, ideally a consecutive observation in a preceding or following sector is advantageous to ensure that the trend is astrophysical instead of some uncorrected systematic effect. Six of the 23 ATLAS targets have more than one sector of observations, as noted in Table \ref{tab:ATLAS32}. L 34-26 was observed in eight sectors and exhibits a clear rotation signal throughout. L 98-59 was observed in the continuous viewing zone (CVZ) during seven sectors and also shows clear rotation, as seen in Figure \ref{fig:L098_059_ALL}. All \textit{TESS} light curves in the 40 sectors for the 23 observed ATLAS stars are shown in Figure \ref{fig:all_TESS}. All but LP 771-95A and L 34-26 had the polynomial included to determine the \texttt{unpopular} fluxes because of the obvious high frequency signal in the data.

\begin{figure}
\epsscale{1}
\includegraphics[angle=90]{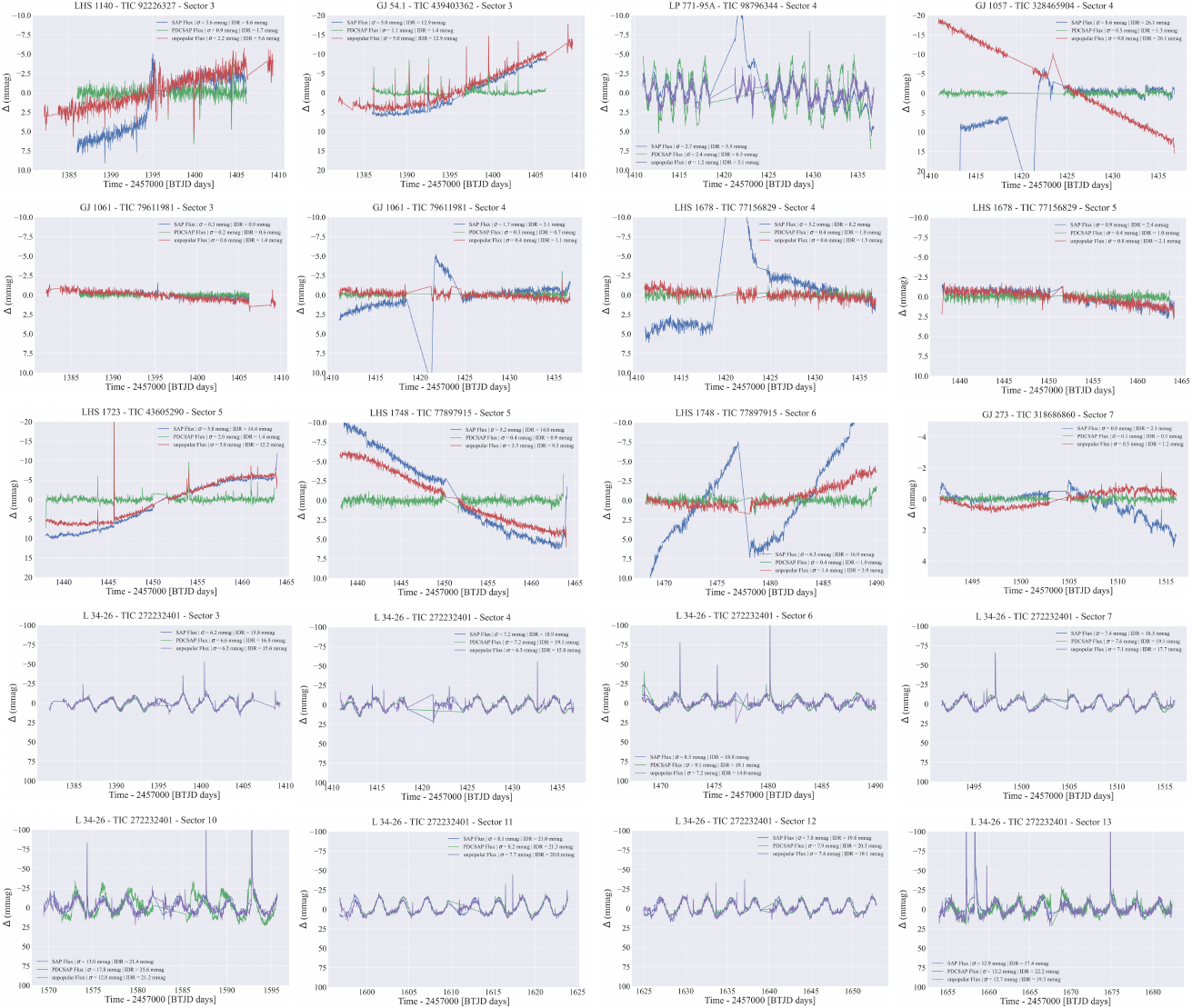}
\caption{From \textit{TESS}, 23 of 32 ATLAS targets were identified whose light curves are shown here and all the variability metrics are reported respectively in Table \ref{tab:ATLAS32}. \textit{TESS} sectors have spatial overlap, and therefore some targets have multiple light curves. The missing targets are a consequence of their location along the ecliptic, which the \textit{TESS} primary mission did not observe. For the \texttt{unpopular} fluxes, LP 771-95A and L 34-26 were reduced without the polynomial (purple) applied while the rest included the polynomial (red). No SPOC fluxes were available for GJ 667C. The light curves are ordered by Right Ascension from left to right and top to bottom. The color code and zero y-axis values are the same as in Figure \ref{fig:RECONS}.}
\label{fig:all_TESS}
\end{figure}

\begin{figure}
\figurenum{11}
\epsscale{1}
\includegraphics[angle=90]{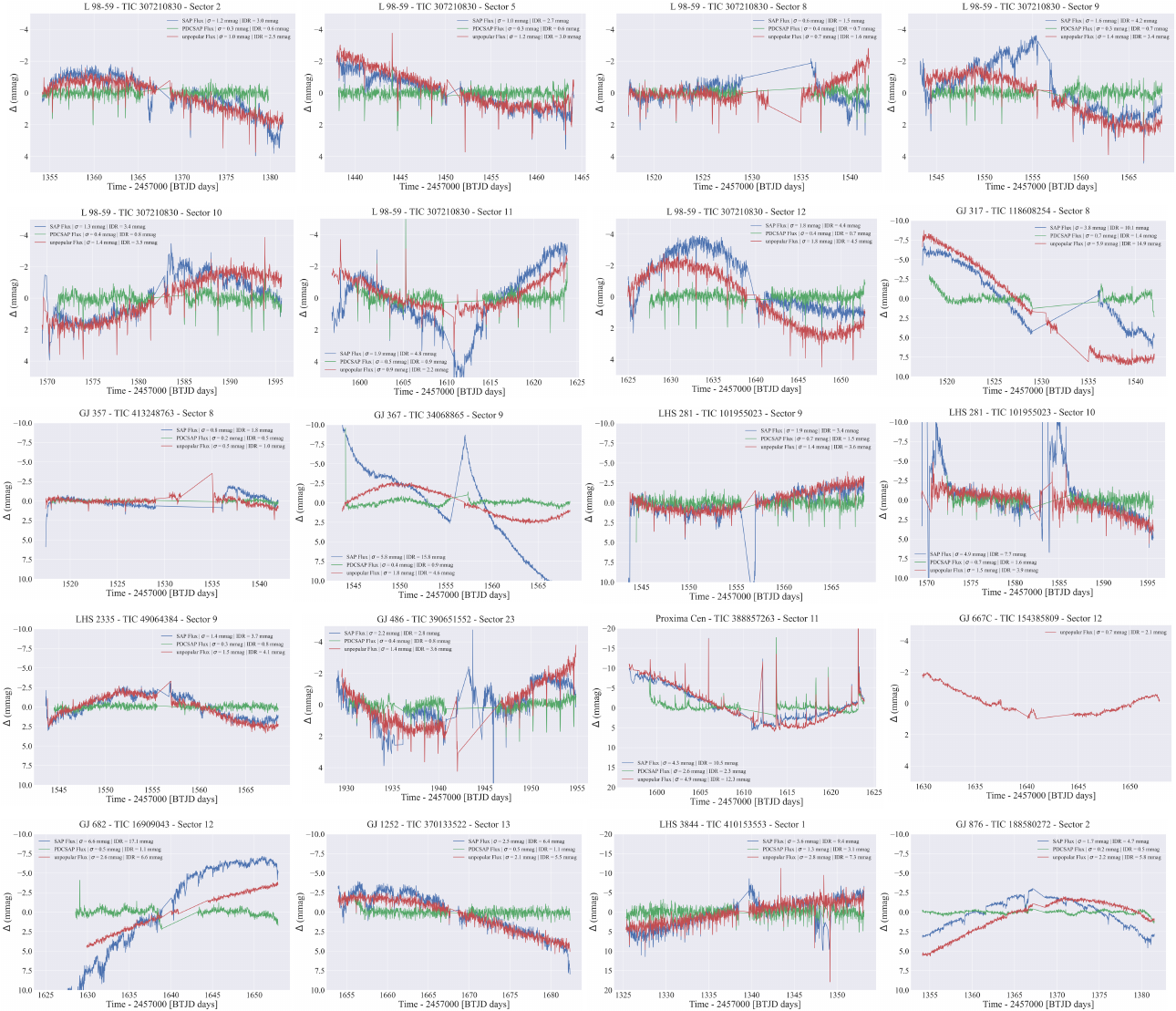}
\caption{\textit{(contd.)}}
\end{figure}

\vfil

\section{Discussion} \label{Discussion}

\subsection{RECONS Long-term Variability and TESS Mid-term Variability} \label{RECONSvsTESS}

\begin{figure}
\epsscale{1.1}
\plotone{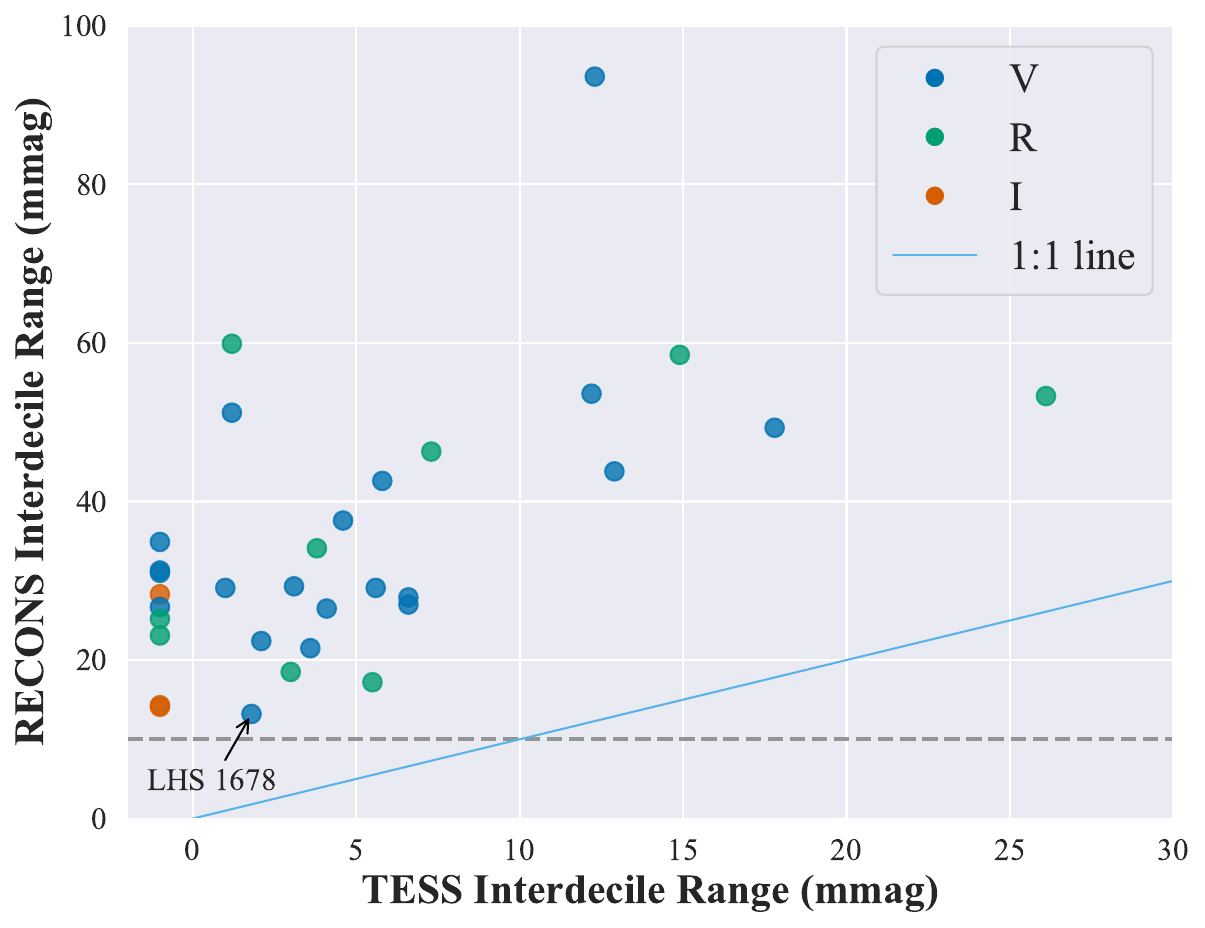}
\caption{The decade+ stellar variability of 32 ATLAS targets observed by RECONS in \textit{VRI} filters (blue, green, and orange, respectively) at the CTIO 0.9 m and the $\sim$month variability observed by \textit{TESS} (averaged over multiple sectors, if available) are shown. The IDR noise floor (horizontal grey dashed line) is set at 10 mmag for RECONS, determined using observations of photometrically stable white dwarfs, whereas the noise floor for \textit{TESS} is $\sim$1 mmag. Nine data points at \textit{TESS} IDR $<$ 0 are targets for which \textit{TESS} data are unavailable due to their locations along the ecliptic and represent 'N.O.' values in Table \ref{tab:ATLAS32}. Note that the y-axis extends to 100 mmag while the x-axis extends to 30 mmag.  The blue line traces 1:1 equal variability on both timescales. As shown in \citet{Kopp2016}, the Sun's Total Solar Irradiance varies by 0.1\%, which would place it in the region near 1 mmag on both axes, assuming the same variations in our filters.}
\label{fig:RECONS_vs_TESS}
\end{figure}

We compare the RECONS and \textit{TESS} variability results in Figure \ref{fig:RECONS_vs_TESS}. From the RECONS results, we find that at longer, multi-decade timescales of years to decades, out of the 32 ATLAS targets, 6 vary $<$ 2\% ($\sim$22 mmag), 25 vary between 2--6\% ($\sim$22--67 mmag), and one, Proxima Centauri, varies by more than 6\%. In contrast, from \textit{TESS} data we find that 17 of the 23 targets show variability $<$ 1\% ($\sim$11 mmag).  Every star falls above the 1:1 line, suggesting that the amplitude of variability is larger over years to decades than at $\sim$month long timescales. However, we note that depending on the physical process of variability, stellar activity can be wavelength dependent. Thus, we caution that this trend is partially a result of the different filters used for the observations because M dwarfs display smaller amplitudes of variability at redder wavelengths \citep{Hosey2015}. The $VRI$ filters used in the RECONS 0.9 m work are bluer than the \textit{TESS} bandpass, which spans the $R$ and $I$ filters and includes even redder wavelengths. The most direct comparison we could make with current data is between the $I$ filter and the \textit{TESS} bandpass, but none of the 23 ATLAS stars with \textit{TESS} observations discussed here were observed with the $I$ filter at the 0.9 m. Overall, this means that given a certain level of spot activity and simultaneous observations, \textit{TESS} would almost certainly show a lower amplitude of variability than our RECONS \textit{V} and \textit{R} filter light curves, despite the same underlying stellar activity.

While no firm claims can be made without careful filter conversions --- simultaneous observations are currently underway in the $VRI$ filters to enable direct comparisons --- it is evident that for some of our targets the long-term variability in RECONS light curves dominates the mid-term variability seen in those same light curves, as well as in the corresponding \textit{TESS} light curves. For example, GJ 1061 shows a mid-term variability of 1.2 mmag with \textit{TESS} and a much larger long-term variability of 59.9 mmag at $R$ from the 0.9 m data. Other targets show less pronounced differences between mid- and long-term variability, such as L 34-26 with 17.8 mmag vs. 49.3 mmag and GJ 1252 with 5.5 mmag vs. 17.2 mmag. Clearly, the detailed balancing between mid-term rotation amplitudes and long-term cycle amplitudes requires continued study to determine when and why one or the other dominates in different kinds of M dwarfs. Such systematic studies are beyond the scope of this paper; here we use the available data to identify the overall least-variable exoplanet-hosting M dwarf systems.

As shown in Figure \ref{fig:RECONS_vs_TESS}, among the 32 ATLAS M dwarfs with planets evaluated here, LHS 1678 is the star that offers the least variable, presumably the most likely habitable, environment within 25 pc. This exohost shows $\leq$ 13.2 mmag of variability at both mid- and long-term timescales. \textit{TESS} data also revealed that GJ 273, GJ 357, and GJ 1061 vary by $\leq$ 1.2 mmag over the $\sim$month long observations. In contrast, at longer timescales they vary by 51.2, 29.1, and 59.9 mmag, respectively. These results indicate that long-term studies are critical because any mid-term studies may not capture the true stellar activity levels. Such studies can help identify exoplanet systems with stable exohosts at different timescales that warrant follow-up observations for exoplanet atmosphere characterizations.

GJ 273, GJ 367, and LHS 1723 show relatively long rotation signals in their \textit{TESS} light curves, but they only have single sector coverage, so it is difficult to confirm their rotation periods. To estimate rotation periods comparable to or longer than a single \textit{TESS} sector baseline of 27.4 days, we will use data from the \textit{TESS} extended mission in future work. This paper only uses the primary mission data, revealing that L 98-59 has by far the clearest rotation signal for a slow rotator at $\sim$40 days with coverage spanning seven sectors, among which five are consecutive. L 34-26 has eight sectors of data, but it is a fast rotator at 2.83 days and the rotation signal is obvious from a single sector alone. Overall, at 26.1 mmag, GJ 1057 is the most variable system over $\sim$month and our closest neighbor, Proxima Centauri, shows the largest variability at long-term timescales, 93.6 mmag at $V$, for which we see only a portion of the rotation period in the \textit{TESS} data from a single sector.

We note that our findings pertain to the present optical variability of these M dwarfs. Unfortunately, the historical variability of these stars remains unknown, and we must acknowledge the potential significance of past variability. Although certain stars like LHS 1678 exhibit low variability amplitude, high activity during the early stages of these M dwarfs may have eroded the atmosphere of the exoplanets in orbit. Follow-up observations for atmospheric characterization of the exoplanets around the ATLAS stars could provide evidence of past stellar variability levels, e.g. the presence of an atmosphere could indicate low stellar activity levels in the past. Furthermore, stars are known to be active when they are younger, but estimating the age of these stars is difficult. We can look to M dwarfs in young clusters for guidance --- cluster studies \citep{Douglas2017,Douglas2019,Curtis2020} have found that M dwarfs exhibit stalled spin down and can stay active for a few Gyr \citep{Pass2022}. Thus, more work is needed to better constrain the ages of these field stars to place them in context of our current activity results.

The trend of higher variability at shorter wavelengths continues beyond the \textit{VRI} optical filters used at the 0.9 m to UV and x-ray wavelengths. At these higher energy levels, we compared our 32 ATLAS stars with the \textit{eROSITA-ROSAT-TESS} sample of 687 M dwarfs in \citet{Magaudda2022} and found only one of the 32 stars to have x-ray data --- LHS 1723 at luminosity levels of log $L_x$ [erg s$^{-1}$] = 28.63 and 28.71 from \textit{eROSITA} and \textit{ROSAT}, respectively. The non-detections for the other 31 stars may imply an absence of detectable x-ray fluxes due to a lack of suitable stellar activity. Further analysis on links between optical and higher energy variability in M dwarfs is reserved for future work with an expanded sample.

\vskip 30pt
\subsection{Exoplanetary Irradiation Levels} \label{Irradiation}

\begin{figure}
\epsscale{1.1}
\plotone{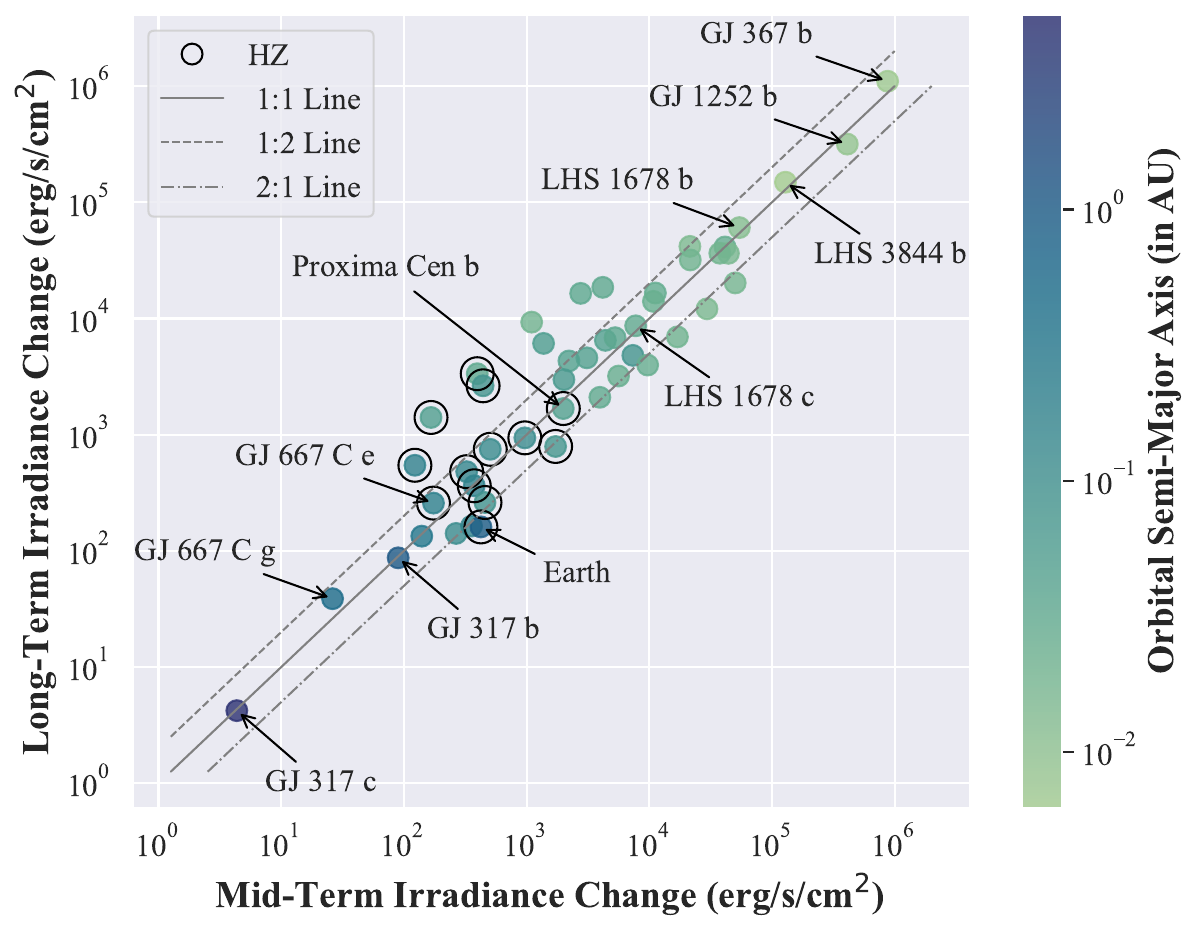}
\caption{The long-term (from RECONS) and mid-term (from \textit{TESS}) changes in irradiance are shown for the 46 exoplanets (excluding Earth) reported to be orbiting the 22 ATLAS stars for which we have both 0.9 m and \textit{TESS} data. Irradiance changes are expressed in $erg/s/cm^2$, calculated for the stars using BT-Settl model stellar fluxes at the planetary orbital distances and using the variability measurements in the respective filters of observation as given in Table \ref{tab:ATLAS32} (described in detail in $\S$\ref{Irradiation}). Points are colored by their reported orbital semi-major axis. A black circle denotes that the exoplanet lies in the HZ at a distance appropriate for liquid water to exist. The grey lines (solid, dashed, dot-dashed) are the 1:1, 1:2, and 2:1 comparison lines between the long-term variability and mid-term variability.}
\label{fig:Irradiation}
\end{figure}

Although we have measured and compared the variability levels between mid-term and long-term timescales for the 23 ATLAS stars, these variations are measured at the stellar photospheres. These M dwarfs were selected to have reported planets, and it is arguably more important to compare how the flux levels change at the planets' locations. The NASA Exoplanet Archive provides a compilation of planetary distances from their host stars, calculated from their respective detection methods as per their discovery papers in addition to stellar effective temperatures and radii. Therefore, here we show the results of calculations of the flux level changes received at the planets given their orbital distances, referred to hereafter as changes in irradiance, with the results shown in Figure \ref{fig:Irradiation}. On this plot, lower stellar fluxes, larger planetary distances, and lower variability levels shift points to the lower left.

To find irradiance levels, we first obtain the relevant stellar surface fluxes. For stars with effective temperatures between 2700 to 3800 K and \textit{log g} = 5 and [Fe/H] = 0, we obtain a grid of synthetic spectra based on the BT-Settl implementation of the PHOENIX model atmosphere code \citep{Allard2003,Allard2011,Allard2012,Allard2013}, from the Spanish Virtual Observatory (SVO) using the SVO Theory Server \footnote{http://svo2.cab.inta-csic.es/theory/newov2/index.php}. We then download the response functions of our respective filter bandpasses (\textit{VRI} at the CTIO 0.9 m and \textit{TESS}) from the SVO Filter Profile Service \citep{Rodrigo2012,Rodrigo2020} and convolve this with our synthetic spectra to get total model fluxes in units of $erg/s/cm^2$ through each bandpass. We then linearly interpolate between these model fluxes as a function of effective temperature, to find the corresponding fluxes for the M dwarfs at their reported effective temperature values from the NASA Exoplanet Archive. Combined with the reported stellar radii, we then find the total flux emitted from the surface of the star in units of $erg/s$ and using the reported planetary orbital semimajor axes, we find the corresponding irradiance levels received by the planets in our respective filter bandpasses in units of $erg/s/cm^2$. We finally multiply this flux at the planet's locations by the variability values measured from the mid-term and long-term results to measure the change in irradiance levels experienced by each planet orbiting an ATLAS star in the filter used for the observations. 

In Figure \ref{fig:Irradiation}, we plot not the irradiance experienced by each of the 46 planets orbiting the 22 ATLAS stars\footnote{The exoplanet reported around L 34-26 has an orbital semi-major axis of more 7000 AU and is not included here. This exoplanet could be a brown dwarf or planet but we do not address it here.}, but the {\it changes} in irradiance which spans more than a factor of 10$^5$ $erg/s/cm^2$. The 1:1 solid grey line is drawn for reference; planets above this line receive larger irradiance changes over the long-term than over the mid-term and conversely for those below this line. Offsets by factors of two are shown with dotted lines. Note that this is a log---log plot and thus some planets lie on the 1:1 line because they have similar long-term and mid-term irradiance changes and any subtle differences get washed out. This implies that the eight planets above the 1:2 line experience drastic long-term irradiance changes during their stellar cycles that are more than twice the mid-term irradiance changes, although recall that the measurements are made in different filters.

For comparison, the Earth's location is shown, assuming 0.1\% flux changes over both mid-term in the \textit{TESS} bandpass and the same hypothetical long-term changes in the $V$ filter \citep{PEVTSOV2023}. The Earth's point is encircled because it is in the HZ of the Sun. Also encircled are exoplanets lying in the HZs around the M dwarfs they orbit. Here we define the HZ to be the extent of the locations where water is anticipated to be in liquid form, on the surface of a planet with an atmosphere, with ranges adopted from \citep{Kopparapu2013} that span recent Venus to early Mars conditions. Using the data from their Figure 7b, we list examples of the HZ distances for early, mid, and late M dwarfs, which cover the spectral types present in the ATLAS sample, along with their corresponding short orbital periods:
\begin{enumerate}
    \item At 0.6 M$_{\odot}$, the HZ is between 0.3 and 0.5 AU, with $P_{orb}$ = 78 to 167 days
    \item At 0.3 M$_{\odot}$, the HZ is between 0.1 and 0.2 AU, with $P_{orb}$ = 21 to 60 days
    \item At 0.1 M$_{\odot}$, the HZ is between 0.03 and 0.06 AU, with $P_{orb}$ = 6 to 17 days
\end{enumerate}

In Figure \ref{fig:Irradiation}, all planets in the HZ cluster together because the distance/emitted flux combinations work in concert to produce temperatures at similar fluxes at the planetary distances. Note that irradiance changes for exoplanets in the HZ span roughly a factor of 30 for both timescales of variability, so exoplanets with liquid water potentially on the surface experience very different flux environments. The planet in the HZ experiencing one of the smallest irradiance changes over both timescales appears to be GJ 667C e. 

Also labeled are three exoplanets with the largest and smallest changes in irradiance, as well as the points representing planets orbiting the least variable star, LHS 1678, the most variable star, Proxima Centauri, and the planet in the HZ, GJ 667C e, that experiences one of the least irradiance changes at both timescales. GJ 317 c experiences the smallest change in irradiance levels over both timescales, but at a distance of 5.23 AU, the planet is much further than the HZ for GJ 317. At the other end of the distribution, GJ 367 b experiences the largest change in irradiance levels because it is only 0.007 AU from GJ 367, which is also among the more variable stars. Although LHS 1678 shows the least stellar variability on both timescales, its planets are found in the top right of the plot because the planets are located much closer to its host star placing it outsize the HZ, and thus results in higher irradiance changes even with slight stellar flux variations. The most variable star in this study is Proxima Centauri, and its planet lies at a distance of 0.049 AU, placing it in the HZ of the host star. 

\section{Systems Worthy of Note} \label{Systems}

ATLAS systems that are nearby, composed of multiple planets or stars, exhibit the least or highest variability in this study, or demonstrate clear long-term rotation signals have been selected for a brief description of their characteristics. These systems of particular interest are listed here, in alphabetical order using the names given in Table \ref{tab:ATLAS32}.

\subsection {GJ 667C}

This is a triple system consisting of two mid-type K dwarfs and an M2.0V star where AB are separated by 0.7\arcsec and AB--C are separated by 33\arcsec, equivalent to a projected separation of $\sim$239 AU; thus, AB do not provide a large proportion of the light falling on the planets. Five exoplanets have been reported to orbit the tertiary star C \citep{Bonfils2013,Anglada2013}, although the existence of GJ 667C e, f and g are subject to debate \citep{Feroz2014,Robertson2014}. The M dwarf is resolved in the RECONS data where we find a variability of 22.4 mmag in \textit{V} over 20 years of observations. However, this star is blended in the \textit{TESS} data, where we find a 2.1 mmag of variability from the combined light of all three components. We find that GJ 667C e, which is fourth most distant planet from the star (if it exists), receives one of the lowest levels of irradiation changes over both timescales compared to other HZ exoplanets and is one of the three exoplanets that are in the HZ of this particular host star.

\subsection{GJ 1061}

With a type of M5.0V, a similar star to Proxima Centauri was discovered by RECONS to be the 20th nearest star system \citep{Henry1997}, located at a distance of only 3.7 pc. It is now reported to have three exoplanets, with two potentially in the HZ \citep{Dreizler2020}. To date, RECONS has observed this target for 23.3 years, finding that it varies by 59.9 mmag in \textit{R}, the second highest long-term variable system in the ATLAS sample, with clear signs of multi-year cyclic variations in the RECONS light curve. This star was observed by \textit{TESS} in Sectors 3 and 4, showing an average variability of only 1.2 mmag, one of the lowest mid-term variables among ATLAS systems. This is the largest variability difference seen between the multi-year and month-long variations among the stars described in this paper.

\subsection{L 98-59}

The L 98-59 system has four confirmed planets \citep{Kostov2019,Demangeon2021}. From the 0.9 m telescope, we have 17.1 years of continuous observations of this target from which we find a variability of 18.5 mmag in \textit{R}. \textit{TESS} has seven sectors of coverage for this target and the location of this system in the CVZ provides a rich dataset from the extended mission that offers even shorter cadences of observations. Five of these sectors were observed consecutively, for which we compute a Lomb-Scargle Periodogram and find a $\sim$40 day rotation period that can be visually verified in the four sequential sectors of \textit{TESS} light curves shown in Figure \ref{fig:L098_059_ALL}. This is half of the reported 80-day rotation period reported in \citet{Cloutier2019}, who used spectroscopic data alone and apparently found a harmonic of the 40-day trend. Over the seven \textit{TESS} sectors, we find an average mid-term variability of 3.0 mmag. This system is among our lowest varying stellar hosts on both timescales and thus is also an excellent candidate to provide favorable environments for its exoplanets. 

\subsection{LHS 1678}

This star is of type M2.0V, is 19.9 pc away, and has two reported exoplanets \citep{Silverstein2021}. The long-term variability from RECONS is low, at only 13.2 mmag in \textit{V}. \textit{TESS} observed this system in Sectors 4 and 5, where the average variability is 1.8 mmag.  This system shows the lowest combination of stellar variability over both the mid- and long-term datasets, making it potentially the most likely habitable environment in the ATLAS sample, and an excellent candidate for follow-up exoplanet atmosphere characterization, given its host star stability.

\subsection{LP 771-95A}

Also known as LTT 1445, this is a triple system consisting of three mid-type M dwarfs (A--BC 7\arcsec, equivalent to a projected separation of 48 AU and BC $<$2\arcsec) at a distance of 6.7 pc. There is one reported exoplanet transiting the primary star A \citep{Winters2019}, which is of type M2.5V. The BC pair is type M3.0VJ and appears to have an orbit coplanar with the orbit of the transiting planet around A.  BC is blended in RECONS images, while all three stars are blended in \textit{TESS}. We find a variability of 29.3 mmag in \textit{V} from the RECONS data for the primary star A (see Figure \ref{fig:RECONS}), and 29.7 mmag for the BC component (not noted in Table \ref{tab:ATLAS32}), making it one of the intermediately variable systems. \textit{TESS} observed this target during Sector 4 and found a variability of 3.1 mmag (see Figure \ref{fig:all_TESS}), but that is for the combined light of all three stars. Rotational modulation can be seen in the \textit{TESS} light curves, which are presumably due to stellar spots on either the B or C component \citep{Winters2019}.

\subsection{Proxima Centauri}

Our closest neighbor is an M5.0V star located at a distance of only 1.3 pc and has one reported exoplanet, located in the HZ \citep{Anglada2016} as highlighted in Figure \ref{fig:Irradiation}. \citet{Wargelin2017} has previously found an 83-day rotation signal with a peak-to-peak 42 mmag amplitude using All Sky Automated Survey photometry in \textit{V}. From the 0.9 m, we find a long-term variability of 93.6 mmag in \textit{V}, the highest in our ATLAS sample. \textit{TESS} observed Proxima during Sector 11, for which we measure a variability of 12.3 mmag, with its slow rotation signature visible in Figure \ref{fig:all_TESS}. Proxima is known to flare, as is evident in the \textit{TESS} light curve. Unfortunately, the high level of variability over both mid- and long-term timescales makes it clear that our next-door neighbor may be less likely to be habitable. Two other potential exoplanets have been reported but are not included in Figure \ref{fig:Irradiation}.

\subsection {2MA2306-0502 (TRAPPIST-1)}

An M7.5V red dwarf located at a distance of 12.5 pc, this star has seven confirmed exoplanets, with a few present in the HZ \citep{Gillon2016,Gillon2017}. RECONS has actively monitored this system for 18.9 years and we find a measured variability of 14.3 mmag in \textit{I}, one of the least variable systems in the ATLAS sample. Our low stellar activity measurements are consistent with other studies \citep{Gillon2017,Roettenbacher2017}. One of the exoplanets within the HZ, TRAPPIST-1 b, has been found to lack an atmosphere using JWST \citep{Lim2023}. The erosion of an exoplanet's atmosphere around a star that currently exhibits low activity levels may point to higher activity levels earlier in the star's life. Unfortunately, \textit{TESS} did not observe this system during its primary mission because of its location in the ecliptic, but it is scheduled to be observed during the extended mission. Although the star is not observed to have high photometric variability in our long-term data, as well as other monitoring campaigns, the JWST observations did reveal at least one spot, posing some challenges for measuring exoplanet transmission spectra \citep{Lim2023}.

\section{Conclusions and Future Work} \label{Conclusion}

This survey finds that the M dwarfs studied here do not vary by more than a few percent at mid-term and long-term timescales at optical wavelengths where they emit much of their flux. Over multi-year to decade timescales, 31 out of the 32 stars in our sample show stellar flux variations $<$ 6\%, while over month-long timescales, 22 out of 23 vary by $<$ 2\%. Note that these levels far exceed the Sun's Total Solar Irradiance fluctuations of $\sim$0.1\% ($\sim$1 mmag) over the 11-year Solar Cycle. It is clear from this study that long-term efforts are key to understanding the behavior of M dwarfs because we typically see (much) higher variability at longer timescales than over mid-term timescales. In this first ATLAS review, LHS 1678 appears to be the best host for potential life-bearing planets and best candidate for atmospheric characterization because its variability levels are $<$ 1.2\% at both mid- and long timescales. However, accounting for stellar flux changes as received at the reported distances of exoplanets orbiting the ATLAS stars, GJ 667C e (if it exists), experiences one of the smallest changes in irradiance at both timescales among the 12 planets orbiting in the HZs of the M dwarfs investigated here. A study of stellar cycles of M dwarfs, some noticeable in the RECONS light curves, will characterize this behavior due to magnetic activity in greater detail (Couperus et al. in prep). For our future work, we will extend our sample to several hundred of the nearest M dwarfs in the southern sky and compare variability of individual stars at various wavelengths by observing them simultaneously with \textit{VRI} filters. We will augment the mid-term data sets with future observations from the \textit{TESS} extended mission and incorporate data for 22 additional exoplanet hosts that have been observed at the 0.9 m. In sum, these efforts will allow us to pursue our quest of following A Trail to Life Around Stars (ATLAS) by revealing the nearest habitable M dwarf systems.

\vskip30pt

This effort has been supported by the NSF through grant AST-2108373. This work has been made possible because of collaborators at the Cerro Tololo Inter-American Observatory (CTIO), and the SMARTS Consortium. This work has made use of data from the European Space Agency (ESA) mission {\it Gaia} (\url{https://www.cosmos.esa.int/gaia}), processed by the {\it Gaia} Data Processing and Analysis Consortium (DPAC, \url{https://www.cosmos.esa.int/web/gaia/dpac/consortium}). Funding for the DPAC has been provided by national institutions, in particular the institutions participating in the {\it Gaia} Multilateral Agreement. Some of the data presented in this paper were obtained from the Mikulski Archive for Space Telescopes (MAST) at the Space Telescope Science Institute. The specific observations analyzed can be accessed via \dataset[https://doi.org/10.17909/0cp4-2j79]{https://doi.org/10.17909/0cp4-2j79} (Sectors 1-26) \citep{MAST_DOI}. STScI is operated by the Association of Universities for Research in Astronomy, Inc., under NASA contract NAS5–26555. Support to MAST for these data is provided by the NASA Office of Space Science via grant NAG5–7584 and by other grants and contracts. This paper includes data collected by the \textit{TESS} mission. Funding for the \textit{TESS} mission is provided by the NASA's Science Mission Directorate. We obtain the data set from the NASA Exoplanet Archive \citep{PSCompPars}\footnote{Accessed on 2023-01-04 at 03:43, returning 5235 rows.}. This dataset or service is made available by the NASA Exoplanet Science Institute at IPAC, which is operated by the California Institute of Technology under contract with the National Aeronautics and Space Administration. This research has made use of the Spanish Virtual Observatory (https://svo.cab.inta-csic.es) project funded by MCIN/AEI/10.13039/501100011033/ through grant PID2020-112949GB-I00. We would also like to thank Soichiro Hattori for his help with the \texttt{unpopular} package. This research made use of Lightkurve, a Python package for Kepler and TESS data analysis \citep{Lightkurve2018}. This work also made use of \texttt{scipy} \citep{Virtanen2020},
\texttt{astropy} \citep{Astropy2013,Astropy2018,Astropy2022},
\texttt{astroquery} \citep{Ginsburg2019},
\texttt{matplotlib} \citep{Hunter2007},
\texttt{numpy} \citep{VanDerWalt2011,Harris2020},
\texttt{tpfplotter} by J. Lillo-Box (publicly available in www.github.com/jlillo/tpfplotter),
\texttt{IRAF} \citep{Tody1986,Tody1993},
and \texttt{SExtractor} \citep{Bertin1996}.

\vfil\eject

\bibliography{references}{}
\bibliographystyle{aasjournal}

\end{document}